\documentclass[pre,twocolumn,aps,floats,floatfix,nofootinbib,citeautoscript,a4paper,showpacs]{revtex4}
\usepackage{epsfig}
\usepackage{graphicx}
\usepackage{amssymb,amsmath}
\usepackage{float}
\usepackage{version}
\usepackage{hyperref}
\usepackage{chemarr}
\usepackage{color}
\usepackage{times}
\bibpunct{[}{]}{,}{n}{}{,}

\newif\ifpdf\ifx\pdfoutput\undefined\pdffalse\else\pdfoutput=1\pdftrue\fi
\newcommand{\pdfgraphics}{\ifpdf\DeclareGraphicsExtensions{.pdf,.jpg,.png}\else\fi}
\newcommand{\fig}{Fig. }
\newcommand{\figs}{Figs. }

\newcommand{\eq}{Eq.}
\newcommand{\eqs}{Eqs.}
\newcommand{\on}{\textrm{on}}
\newcommand{\off}{\textrm{off}}
\newcommand{\sce}{\textrm{SCE}}
\newcommand{\sse}{\textrm{SSE}}
\newcommand{\pon}{p_{\textrm{on}}}
\newcommand{\hon}{h_{\textrm{on}}}
\newcommand{\hont}{{\tilde h}_{\textrm{on}}}
\newcommand{\Gon}{G_{\textrm{on}}}

\newcommand{\kon}{k^{\textrm{on}}}

\newcommand{\poff}{p_{\textrm{off}}}
\newcommand{\Goff}{G_{\textrm{off}}}

\newcommand{\koff}{k^{\textrm{off}}}
\newcommand{\Prob}{{\textrm{Prob}}}

\newcommand{\ks}{k^{s}}
\newcommand{\kms}{k^{1-s}}
\newcommand{\ps}{p_{s}}
\newcommand{\pms}{p_{1-s}}

\newcommand{\hoff}{h_{\textrm{off}}}
\newcommand{\hofft}{{\tilde h}_{\textrm{off}}}
\newcommand{\htd}{{\tilde h}}
\newcommand{\tW}{{\widetilde W}}

\newcommand{\dd}{{d}}
\newcommand{\s}{{s}}

\newcommand{\p}{\partial}
\newcommand{\cL}{{\cal L}}

\providecommand{\avg}[1]{\left \langle #1 \right \rangle}
\graphicspath{{figures/},{figures/pdf/},{figures/eps/}}

\begin{document}
\pdfgraphics

\title{Statistical physics of a model binary genetic switch with linear
feedback}

\author{Paolo Visco}

\affiliation{SUPA, School of Physics and Astronomy, The University of
Edinburgh, James Clerk Maxwell Building, The King's Buildings,
Mayfield Road, Edinburgh EH9 3JZ, UK}

\author{Rosalind J. Allen}

\affiliation{SUPA, School of Physics and Astronomy, The University of
Edinburgh, James Clerk Maxwell Building, The King's Buildings,
Mayfield Road, Edinburgh EH9 3JZ, UK}

\author{Martin R. Evans}

\affiliation{SUPA, School of Physics and Astronomy, The University of
Edinburgh, James Clerk Maxwell Building, The King's Buildings,
Mayfield Road, Edinburgh EH9 3JZ, UK}

\date{\today}
\begin{abstract}
We study the statistical properties of a simple genetic regulatory
network that provides heterogeneity within a population of cells. This
network consists of a binary genetic switch in which stochastic
flipping between the two switch states is mediated by a ``flipping''
enzyme.  Feedback between the switch state and the flipping rate is
provided by a linear feedback mechanism: the flipping enzyme is only
produced in the on switch state and the switching rate depends
linearly on the copy number of the enzyme. This work generalises the
model of [{\it Phys. Rev. Lett.}, {\bf 101}, 118104] to a broader
class of linear feedback systems. We present a complete analytical
solution for the steady-state statistics of the number of enzyme
molecules in the on and off states, for the general case where the
enzyme can mediate flipping in either direction. For this general case
we also solve for the flip time distribution, making a connection to
first passage and persistence problems in statistical physics. We show
that the statistics of the model are non-Poissonian, leading to a peak
in the flip time distribution. The occurrence of such a peak is
analysed as a function of the parameter space. We present a new
relation between the flip time distributions measured for two relevant
choices of initial condition. We also introduce a new correlation
measure to show that this model can exhibit long-lived temporal
correlations, thus providing a primitive form of cellular memory.
Motivated by DNA replication as well as by evolutionary mechanisms
involving gene duplication, we study the case of two switches in the
same cell. This results in correlations between the two switches;
these can either positive or negative depending on the parameter
regime.
\end{abstract}


\maketitle

\section{Introduction}

Populations of biological cells frequently show stochastic switching
between alternative phenotypic states. This phenomenon is particularly
well-studied in bacteria and bacteriophages, where it is known as
phase variation \cite{woude2004}. Phase variation often affects cell
surface features, and its evolutionary advantages are believed to
involve evading attack from host defense systems (e.g. the immune
system) and/or ``bet-hedging'' against sudden catastrophes which may
wipe out a particular phenotypic type. Switching between different
phenotypic states is controlled by an underlying genetic regulatory
network, which randomly flips between alternative patterns of gene
expression. Several different types of genetic network are known to
control phase variation---these include DNA inversion switches, DNA
methylation switches and slipped strand mispairing mechanisms
\cite{woude2004,blomfield2001,lim2007}.

In this paper, we study a simple model for a genetic network that
allows switching between two alternative states of gene expression.
Its key feature is that it includes a linear feedback mechanism
between the switch state and the flipping rate. When the switch is
active, an enzyme is produced and the rate of switching is linearly
proportional to the copy number of this enzyme.  The statistical
properties of this model are made non-trivial by this feedback,
leading, among other things, to non-Poissonian behaviour that may be
of advantage to cells in surviving in certain dynamical environments.
Our model is very generic and does not aim to describe any specific
molecular mechanism in detail, but rather to determine in a general
way the consequences of the linear feedback for the switching
statistics. Motivated by the fact that cells often contain multiple
copies of a particular genetic regulatory element, due to DNA
replication or DNA duplication events during evolution, we also
consider the case of two identical switches in the same cell. We find
that the two copies of the switch are coupled and may exhibit
interesting and potentially important correlations or
anti-correlations. Our model switch is fundamentally different from
bistable gene networks that have been the subject of previous
theoretical interest. In fact, as we shall show, our switch is not
bistable but is intrinsically unstable in each of its two states.

Before discussing our model in detail, we provide a brief overview of
the basic biology of genetic networks and summarise some previously
considered models for genetic switches. Genetic networks are
interacting, many-component systems of genes, RNA and proteins, that
control the functions of living cells. Genes are stretches of DNA
($\sim$1000 base pairs long in bacteria), whose sequences encode
particular protein molecules. To produce a protein molecule, the
enzyme complex RNA polymerase copies the gene sequence into a
messenger RNA (mRNA) molecule. This is known as transcription. The
mRNA is then translated (by a ribosome enzyme complex) into an amino
acid chain which folds to form the functional protein molecule. The
production of a specific set of proteins from their genes ultimately
determines the phenotypic behaviour of the cell. Phenotypic behaviour
can thus be controlled by turning genes on and off. Regulation of
transcription (production of mRNA) is one important way of achieving
this. Transcription is controlled by the binding of proteins known as
transcription factors to specific DNA sequences, known as operators,
usually situated at the beginning of the gene sequence. These
transcription factors may be activators (which enhance the
transcription of the gene they regulate) or repressors (which repress
transcription, often by preventing RNA polymerase binding). A given
gene may encode a transcription factor that regulates itself or other
genes, leading to complex networks of transcriptional interactions
between genes.

There has been much recent interest among both physical scientists and
biologists in deconstructing complex genetic networks into modular
units \cite{alon}, and in seeking to understand their statistical
properties using theory and simulation~\cite{sneppen,alonbook}. Of
particular interest is the fact that genetic networks are
intrinsically stochastic, due to the small numbers of molecules
involved in gene expression \cite{elowitz,swain}. This can give rise
to heterogeneity in populations of genetically and environmentally
identical cells \cite{elowitz}. For some genetic networks, this
heterogeneity is ``all-or-nothing'': the population splits into two
distinct sub-populations, with different states of gene
expression. Such networks are known as bistable genetic switches: they
have two possible long-time states, corresponding to alternative
phenotypic states. Well-known examples are the switch controlling the
transition from the lysogenic to lytic states in bacteriophage
$\lambda$ \cite{lambdap,OKSCA05}, and the lactose utilisation network
of the bacterium {\em{Escherichia coli}} \cite{ozbudak}.  Several
simple mechanisms for achieving bistability have been studied,
including pairs of mutually repressing genes \cite{CA00,gardner},
positive feedback loops \cite{farrell} and mixed feedback loops
\cite{FH05}. Such bistable genetic networks can allow long-lived and
binary responses to short-lived signals---for example, when a cell is
triggered by a transient signal to commit to a particular
developmental pathway.

Theoretical treatments of bistable genetic networks usually consider
the dynamics of the copy number (or concentration) of the regulatory
proteins involved. This affects the activation state of the genes,
which in turn influences the rate of protein production. The
macroscopic rate equation approach \cite{Keller95} provides a
deterministic (mean-field) description of the dynamics that ignores
fluctuations in protein copy number or gene expression state.  This
approach, applied to a switch with two mutually repressing genes, has
shown that co-operative binding of regulatory proteins is an important
factor in generating bistability \cite{CA00}. Other studies have
shown, however, that bistability can be achieved even when the
deterministic equations have only one solution, due to stochasticity
and fluctuations in protein numbers \cite{LLBB06,artyomov}. An
alternative approach is to study the dynamics of stochastic flipping
between two stable states using stochastic simulations
\cite{warren2004,WtW05,morelli2008}, by numerically integrating the
master equation \cite{LLBB07}, or by path integral-type approaches
\cite{aurell}.  This dynamical problem bears some resemblance to the
Kramers problem of escape from a free energy minimum
\cite{Bialek00,KE01}, and one expects on general grounds that the
typical time spent in one of the bistable states should be
exponentially large in the typical number of proteins present in the
state. This has been confirmed, at least for cooperative toggle
switches formed of mutually repressing genes
\cite{warren2004,WtW05}. From the perspective of statistical physics,
interesting questions arise concerning the distribution of escape
times and the connection to first passage properties of stochastic
processes.

In this paper, however, we are concerned with an intrinsically
different situation from these bistable genetic networks. The
molecular mechanisms controlling microbial phase variation typically
involve a binary element that can be in either of two states. For
example, this may be a short fragment of DNA that can be inserted into
the chromosome in either of two orientations, a repeated DNA sequence
that can be altered in its number of repeats, or a DNA sequence that
can have two alternative patterns of methylation \cite{woude2004}. The
flipping of this element between its two states is stochastic, with a
flipping rate that is controlled by various regulatory proteins, the
activity of which may be influenced by environmental factors. We shall
consider the case where a feedback exists between the switch state and
the flipping rate. This is particularly interesting from a statistical
physics point of view because it leads to non-Poissonian switching
behaviour, as we shall show. Our work has been motivated by several
examples. The {\it fim} system in uropathogenic strains of the
bacterium {\em{E. coli}} controls the production of Type 1 fimbriae
(or pili), which are ``hairs'' on the surface of the
bacterium. Individual cells switch stochastically between ``on'' and
``off'' states of fimbrial production
\cite{woude2004,wolf2002,CB07,gally1993}.  The key feature of the {\it
fim} switch is a short piece of DNA that can be inserted into the
bacterial DNA in two possible orientations. Because this piece of DNA
contains the operator sequence for the proteins that make up the
fimbriae, in one orientation, the fimbrial genes are transcribed and
fimbriae are produced (the ``on'' state) and in the other orientation,
the fimbrial genes are not active and no fimbriae are produced (the
``off'' state). The inversion of this DNA element is mediated by
recombinase enzymes. Feedback between the switch state and the switch
flipping rate arises because the FimE recombinase (which flips the
switch in the on to off direction), is produced more strongly in the
on switch state than in the off state. This phenomenon is known as
orientational control \cite{kulasekara1999,joyce2002,hinde2005}. The
production of a second type of fimbriae in uropathogenic
{\em{E. coli}}, Pap pili, also phase varies, and is controlled by a
DNA methylation switch \cite{woude2004,blomfield2001,low1987}. Here,
the operator region for the genes encoding the Pap pili can be in two
states, in which the DNA is chemically modified (methylated) at
different sites, and different binding sites are occupied by the
regulatory protein Lrp. Switching in this system is facilitated by the
PapI protein, which helps Lrp to bind \cite{Nou1995}. Feedback between
the switch state and the flipping rate arises because the production
of PapI itself is activated by the protein PapB, which is only
produced in the ``on'' state
\cite{Goransson1989,woude2004,blomfield2001}.

A common feature of the above examples is the existence of a feedback
mechanism: in the {\it fim} system this occurs through orientational
control, and in the {\it{pap}} system, through activation of the
{\em{papI}} gene by PapB. In this paper, we aim to study the role of
such feedback within a simple, generic model of a binary genetic
switch.  We shall assume that the feedback is linear, and we thus term
our model a ``linear feedback switch''.  In a recent publication
\cite{VAE08}, we introduced a simple mathematical model of a DNA
inversion genetic switch with orientational control, which was
inspired by the {\it fim} system. Our model reduces to the dynamics of
the number of molecules of a ``flipping enzyme'' $R$, which mediates
switch flipping, along with a binary switch state. Enzyme $R$ is
produced only in the on switch state.  As the copy number of $R$
increases, the on to off flipping rate of the switch increases and
this results in a non-Poissonian flipping process with a peak in the
lifetime of the on state.  The model is linear in the sense that
the rate at which the switch is turned off is a linear function of the
number of enzymes $R$ which it produces.  In our previous work
\cite{VAE08}, we imagined enzyme $R$ to be a DNA recombinase, and the
two switch states to correspond to different DNA orientations, in
analogy with the {\em{fim}} system. However, the same model could be
used to describe a range of molecular mechanisms for binary switch
flipping with feedback between the switch state and flipping rate, and
can thus be considered a generic model of a genetic switch with linear
feedback.

In our recent work \cite{VAE08}, we obtained exact analytical
expressions for the steady state enzyme copy number for our model
switch with linear feedback, in the particular case where the flipping
enzyme switches only in the on to off direction (this being the
relevant case for {\it{fim}}).  We also calculated the flip time
distribution for this model analytically. Conceptually, such a
calculation is reminiscent of the study of persistence in statistical
physics \cite{Majumdar99} where, for example, one asks about the
probability that a spin in an Ising system has not flipped up to some
time \cite{DBG94}. For the flip time distribution, we introduced
different measurement ensembles according to whether one starts the
time measurement from a flip event (the Switch Change Ensemble) or
from a randomly selected time (the Steady State Ensemble). In the
present paper, we extend this work to present the full solution of the
general case of the model and extend our study of its persistence
properties. The introduction of a rate for the enzyme mediated off to
on flipping ($\koff_3$) has most significant effects on the flip time
distributions $F(T)$, as illustrated in \figs \ref{fig:diagram} and
\ref{fig:diagramk3off0} where we show the parameter range over which a
peak is found in $F(T)$ for zero and non-zero $\koff_3$. We also prove
an important relation between the two measurement ensembles defined in
\cite{VAE08} and use it to show that a peak in the flip time
distribution only occurs in the Switch Change Ensemble and not in the
Steady State Ensemble. We find that the non-Poissonian behaviour of
this model switch leads to interesting two-time autocorrelation
functions.  We also study the case where we have two copies of the
switch in the same cell and find that these two copies may be
correlated or anticorrelated, depending on the parameters of the
model, with potentially interesting biological implications.

The paper is structured as follows.  In section II we define the
model, describe its phenomenology, and show that a ``mean-field'',
deterministic version of the model has only one steady state solution.
In section III we present the general solution for the steady state
statistics and in section IV we study first passage-time properties of
the switch; technical calculations are left to the appendices.  In
section V we consider two coupled model switches and we present our
conclusions in section VI.

\section{The model}

\begin{figure}
\includegraphics[width=\columnwidth,clip=true]{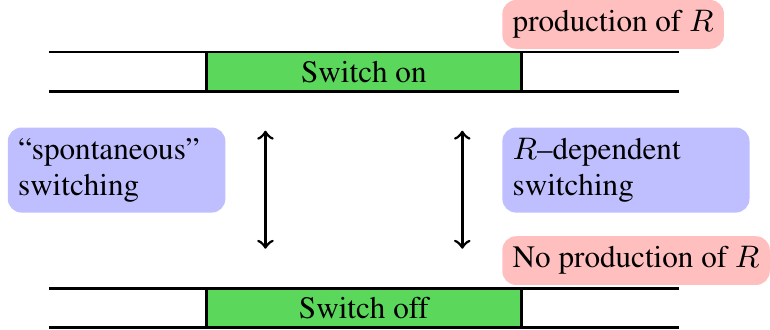}
\caption{\label{fig:sketch}(colour online) A schematic illustration of
the model DNA inversion switch. }
\end{figure}

We consider a model system with a flipping enzyme $R$ and a binary
switch $S$, which can be either on or off (denoted respectively as
$S_\on$ and $S_\off$).  Enzyme $R$ is produced (at rate $k_2$) only
when the switch is in the on state, and is degraded at a constant rate
$k_1$, regardless of the switch state. This represents protein removal
from the cell by dilution on cell growth and division, as well as
specific degradation pathways. Switch flipping is assumed to be a
single step process, which can either be catalysed by enzyme $R$, with
rate constants $\kon_3$ and $\koff_3$ and linear dependence on the
number of molecules of $R$, or can happen ``spontaneously'', with
rates $\kon_4$ and $\koff_4$. We imagine that the ``spontaneous''
switching process may in fact be catalysed by some other enzyme whose
concentration remains constant and which is therefore not modelled
explicitly here. Our model, which is shown schematically in \fig
\ref{fig:sketch}, is defined by the following set of biochemical
reactions:
\begin{subequations}
\label{eq:react}
\begin{align}
\label{eq:reacta}
R & \stackrel{k_1}{\longrightarrow}\emptyset & S_{\on} &
\stackrel{k_2}{\longrightarrow} S_{\on}+R \\
\label{eq:reactb}
S_{\on} + R & \xrightleftharpoons[\koff_3]{\kon_3} S_{\off} + R &
S_{\on} & \xrightleftharpoons[\koff_4]{\kon_4} S_{\off}\,\,.
\end{align}
\end{subequations}

\subsection{Phenomenology}\label{sec:phen}
\begin{figure*}
\includegraphics[width=\textwidth,clip=true]{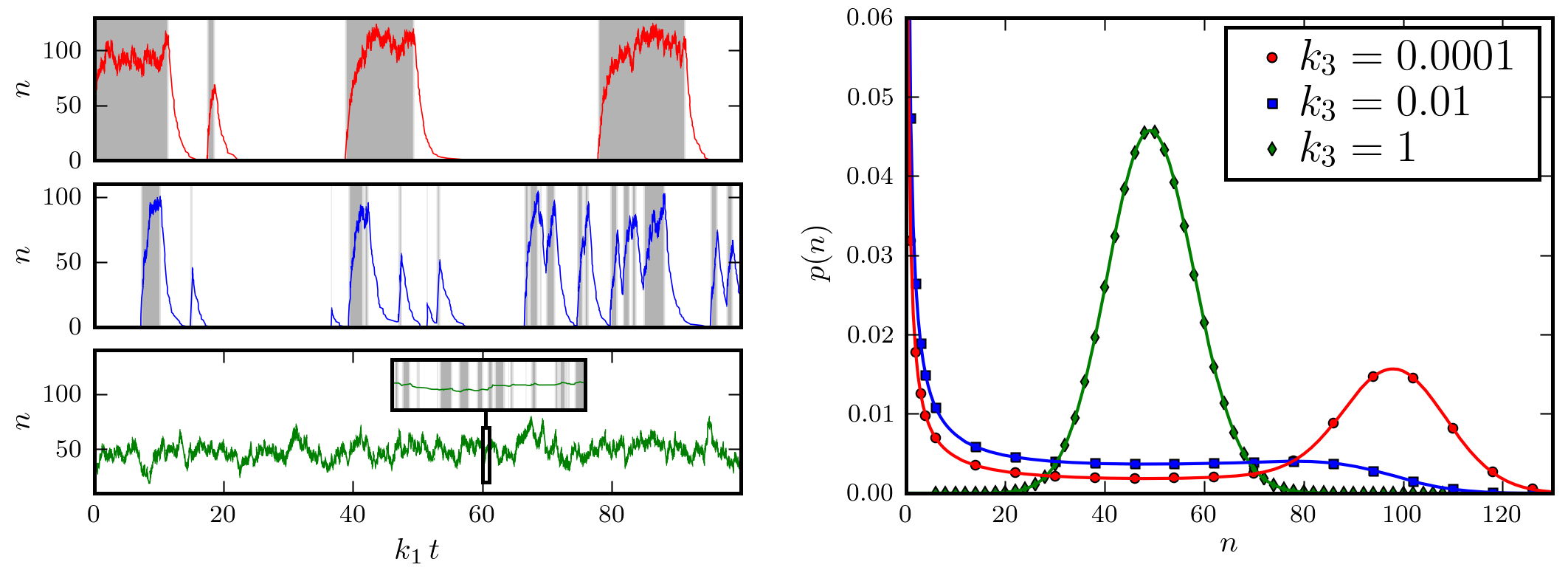}
\caption{\label{fig:sampletraj_k3}(colour online) \textsc{Left}:
Typical trajectories of the system when $\kon_3=\koff_3=k_3$ is
increased (from top to bottom $k_3=0.0001$, 0.01 and 1). The other
parameters are $k_1=1$, $k_2=100$ and $\kon_4=\koff_4=k_4=0.1$. Grey
shading denotes periods in which the switch is in the on state, and
the solid lines denote the number of enzyme molecules, plotted against
time.  In the bottom panel, the switch flips so fast that the grey
shading is only shown in the inset where the trajectory from $k_1 t =
[60,61]$ is shown in detail.  \textsc{Right}: Probability distribution
functions for the number $n$ of $R$ molecules, for parameter values
corresponding to the trajectories shown in the left panels.  The
symbols are the result of numerical simulations (see text for
details).  The full curves plot the analytical results
Eqs. (\ref{eq:ponsol}) and (\ref{eq:poffsol}), which are in perfect
agreement with the simulations.  }

\end{figure*}

We notice that there are two physically relevant and coupled
timescales for our model switch: the timescale associated with changes
in the number of $R$ molecules (dictated by the production and decay
rates $k_1$ and $k_2$), and that associated with the flipping of the
switch (dictated by $k_3$, $k_4$ and the $R$ concentration).

We first consider the case where the timescale for $R$
production/decay is much faster than the switch flipping
timescale. The top left panel of \fig \ref{fig:sampletraj_k3} shows a
typical dynamical trajectory for parameters in this regime.  Here, we
plot the number $n$ of $R$ molecules, together with the switch state,
against time. This result was obtained by stochastic simulation of
reaction set (\ref{eq:react}) using the Gillespie
algorithm~\cite{bortz75,gillespie1976}.  This algorithm generates a
continuous time Markov process which is exactly described by the
master equation (\ref{eq:master}).  For a given switch state, the
number $n$ of molecules of $R$ varies according to reactions
(\ref{eq:reacta}). When the switch is in the on state, $n$ grows
towards a plateau value, and when the switch is in the off state, $n$
decreases exponentially towards $n=0$.  The time evolution of $n$ can
thus be seen as a sequence of relaxations towards two different
asymptotic steady states, which depend on the switch position. To
better understand this limiting case, we can make the assumption that
the number of $R$ molecules evolves deterministically for a given
switch state. We can then write down deterministic rate equations
corresponding to the reaction scheme (\ref{eq:react}).  These
equations are first order differential equations for $\rho$, the mean
concentration of the enzyme.  When the switch is on, the rate equation
reads
\begin{equation}
\frac{d \rho}{d t} = -k_1 \rho + k_2
\end{equation}
with solution
\begin{equation}
\rho(t) = \rho(0) e^{-k_1t} + \frac{k_2}{k_1}\left[ 1- {e}^{-k_1 t} \right]\;.
\end{equation}
Thus  the plateau density in the on state is
given by the
ratio 
\begin{equation}
\rho_{\rm on}  = k_2/k_1\;,
\label{rhoon}
\end{equation}
and the timescale for relaxation to this density is given by $k_1$,
the rate of degradation of $R_1$. When the switch is in the off state,
the rate equation for $\rho$ reads instead
\begin{equation}
\frac{d \rho}{d t} = -k_1 \rho
\end{equation}
and one simply has exponential decay to $\rho=0$ with decay time
$k_1$. In this parameter regime, switch flipping typically happens
when the number of molecules of $R$ has already reached the steady
state (as in the top left panel of \fig
\ref{fig:sampletraj_k3}). Thus, the on to off switching timescale is
given by $1/(\rho_\on \kon_3 + \kon_4)$, where $\rho_{\rm on}$ is the
plateau concentration of flipping enzyme when the switch is in the on
state, given by Eq.(\ref{rhoon}).  Since the corresponding plateau
concentration in the off switch state is zero, the off to on switch
flipping timescale is simply given by $1/\koff_4$.

We now consider the opposite scenario, in which switching occurs on a
much shorter timescale than relaxation of the enzyme copy number. A
typical trajectory for this case is shown in the bottom left panel of
\fig \ref{fig:sampletraj_k3}. Here, switching reactions dominate the
dynamics of the model, and the dynamics of the enzyme copy number
follows a standard birth-death process, with an effective birth rate
given by the enzyme production rate in the on state multiplied by the
fraction of time spent in the on state. A more quantitative
account for these behaviours is provided later on, in \ref{sec:pss}.

For parameter values between these two extremes, where the timescales
for switch flipping and enzyme number relaxation are similar, it is
more difficult to provide intuitive insights into the behaviour of the
model. A typical trajectory for this case is given in the middle left
panel of \fig \ref{fig:sampletraj_k3}. Here, we have set the on to off
and off to on switching rates to be identical: $\kon_3 = \koff_3$ and
$\kon_4 = \koff_4$. We notice that typically, less time is spent in
the on state than in the off state. As soon as the switch flips into
the on state, the number of $R$ molecules starts increasing and the on
to off flip rate begins to increase. Consequently, the number of $R$
molecules rarely reaches its plateau value before the switch flips
back into the off state.

\begin{figure*}
\includegraphics[width=\textwidth,clip=true]{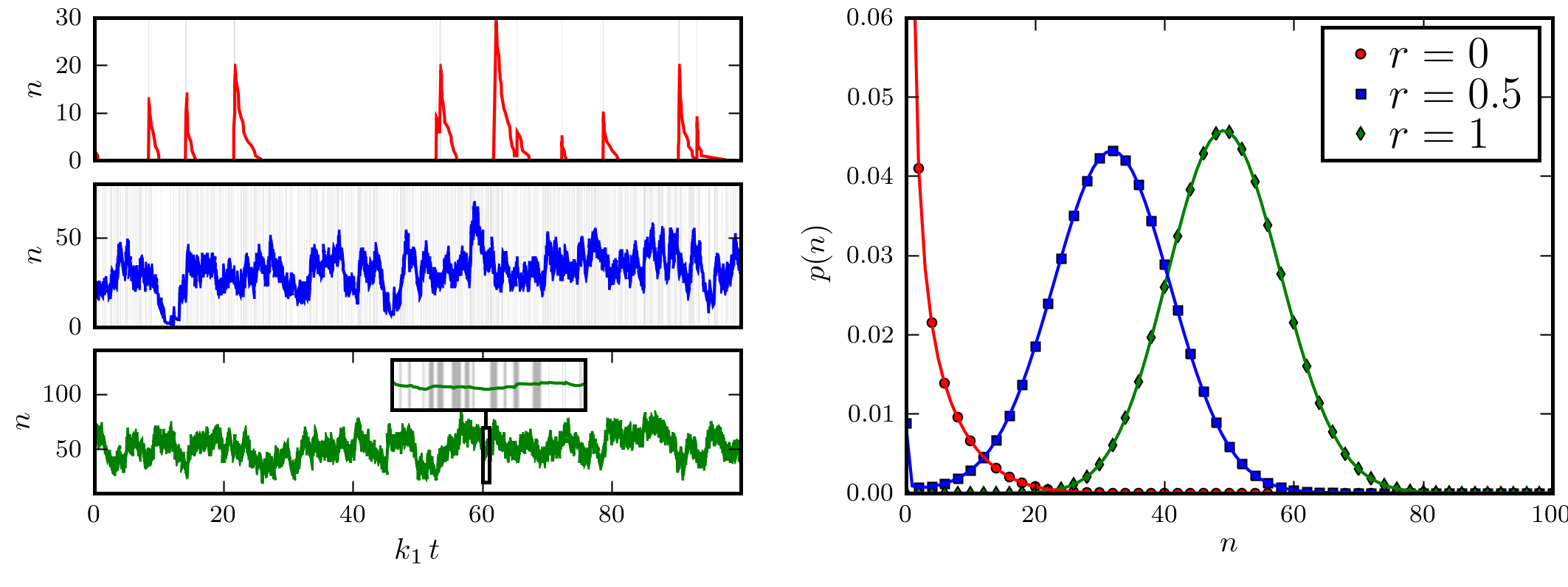}
\caption{\label{fig:sampletraj_k3off}(colour online) \textsc{Left}:
Typical trajectories of the system when $r=\koff_3/\kon_3=$ is
increased (from top to bottom $r=0$, 0.5 and 1). The other parameters
are $k_1=1$, $k_2=100$, $\kon_3=1$ and $\kon_4=\koff_4=k_4=0.1$. Grey
shading denotes periods in which the switch is in the on state, and
the solid lines denote the number of enzyme molecules, plotted against
time.  In the bottom panel, the switch flips so fast that the grey
shading is only shown in the inset where the trajectory from $k_1 t =
[60,61]$ is shown in detail.  \textsc{Right}: Probability distribution
functions for the number $n$ of $R$ molecules, for parameter values
corresponding to the trajectories shown in the left panels.  The
symbols are the result of numerical simulations (see text for
details).  The full curves plot the analytical results
Eqs. (\ref{eq:ponsol}) and (\ref{eq:poffsol}), which are in perfect
agreement with the simulations.  }
\end{figure*}

To illustrate the effects of including the parameter $\koff_3$,
we also show trajectories for different values of the ratio
$r=\koff_3/\kon_3$ in \fig \ref{fig:sampletraj_k3off}, for fixed
$\kon_3$. For small $r$, the amount of enzyme decays to zero in the
off state before the next off-to-on flipping event resulting in bursts
of enzyme production. In contrast, when $r$ is $O(1)$, flipping is
rapid in both directions so that $p(n)$ is peaked at intermediate $n$.

\subsection{Mean-field equations}
\begin{figure*}
\includegraphics[width=\textwidth,clip=true]{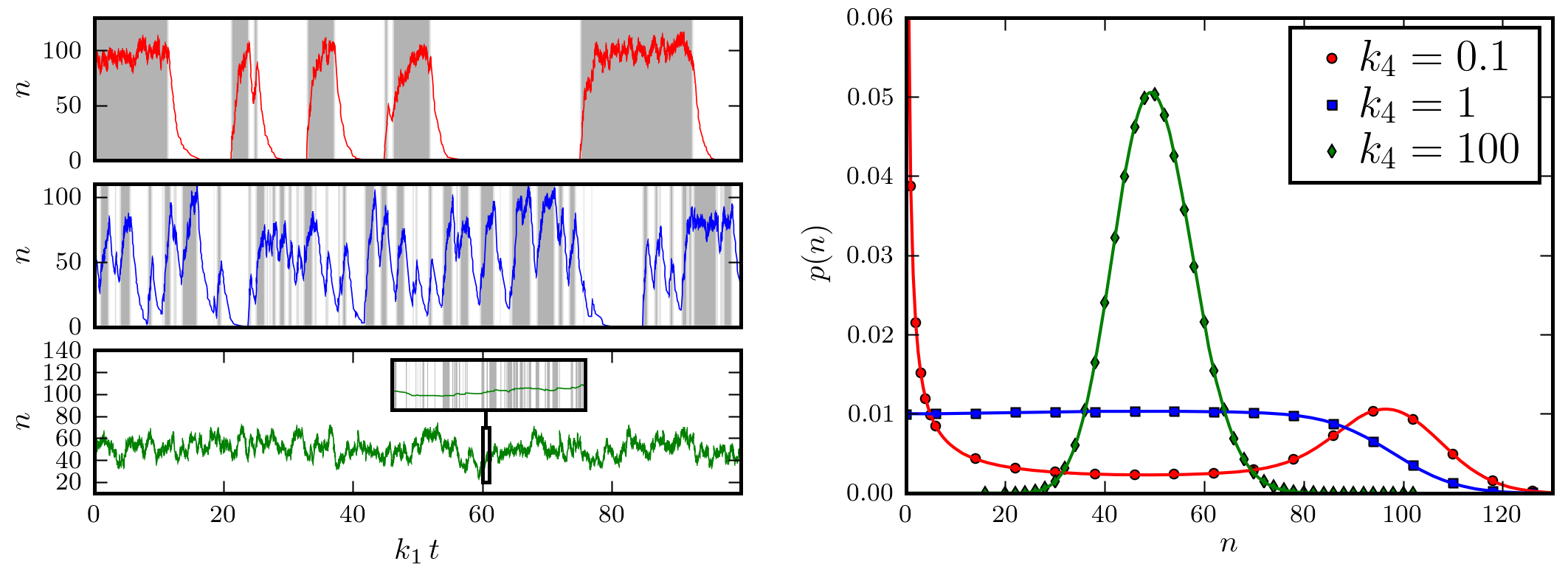}
\caption{\label{fig:sampletraj_k4}(colour online) \textsc{Left}:
Typical trajectories of the system when $\kon_4=\koff_4=k_4$ is
increased (from top to bottom $k_4=0.1$, 1 and 100. Other parameters
are $k_1=1$, $k_2=100$ and $\kon_3=\koff_3=k_3=0.001$. In each panel
the grey shading denotes that the switch is on and the line plots the
number of enzymes against time.  In the third panel the grey shading
is only shown in the inset where the trajectory from $k_1 t = [60,61]$
is detailed.  \textsc{Right}: Probability distribution functions of
the number of $R_1$ molecules in the cell for parameter values
corresponding to the trajectories shown in the left panels.  The
symbols are the result of numerical simulations (see text for
details).  The full curves plot the analytical results
Eqs. (\ref{eq:ponsol}) and (\ref{eq:poffsol}) and pass perfectly
through the simulation points.  }
\end{figure*}

To explore how the switching behaviour of our model arises, we can
write down mean-field, deterministic rate equations corresponding to
the full reaction scheme (\ref{eq:react}). These equations describe
the time evolution of the mean concentration $\rho (t)$ of $R$
molecules and the probabilities $Q_\on (t)$ and $Q_\off (t)$ of the
switch being in the on and off states.  These equations implicitly
assume that the mean enzyme concentration $\rho$ is completely
decoupled from the state of the switch.  Thus correlations between the
concentration $\rho$ and the switch state are ignored and the
equations furnish a mean-field approximation for the switch.  As we
now show, this crude type of mean-field description is insufficient to
describe the stochastic dynamics of the switch, except in the limit of
high flipping rate. Noting that $Q_\on(t) + Q_\off(t)=1$, the
mean-field equations read:
\begin{subequations}
\begin{equation}
\frac{d \rho(t)}{d t}=k_2 Q_\on(t) - k_1 \rho(t) \,\,,
\end{equation}
\begin{multline}
\frac{d Q_\on(t)}{d t}= (\koff_4 + \rho(t) \koff_3) (1-Q_\on(t)) \\-
(\kon_4 + \rho(t) \kon_3) Q_\on(t)\,\,.
\end{multline}
\end{subequations}
The above equations have two sets of possible solutions for the
steady--state values of $\rho$ and $Q_\on$, but only one has a positive
value of $\rho$, and is therefore physically meaningful. The result
is:
\begin{equation}
\rho=\frac{ \rho_{\rm on} {\koff_3}-
   ({\koff_4}+{\kon_4})+\sqrt{\Delta}}{2 ({\koff_3}+{\kon_3})} \,\,,
\label{rho}
\end{equation}
where
\begin{equation}
\Delta=( \rho_{\rm on}{\koff_3}-({\koff_4}+{\kon_4}))^2+4
   \rho_{\rm on} {\koff_4}({\koff_3}+{\kon_3})
   \,\,,
\end{equation}

and
\begin{equation}
Q_\on= \rho/\rho_{\on} \,\,.
\end{equation}
The most interesting conclusion to be drawn from this mean-field
analysis is that there is only one physically meaningful solution. In
this solution, the enzyme concentration $\rho$ is less than the
plateau value in the on state [$\rho_{\rm on}$ of
Eq.(\ref{rhoon})]. Thus reaction scheme (\ref{eq:react}) does not have
an underlying bistability. The two states of our stochastic switch
evident in Figures \ref{fig:sampletraj_k3} and \ref{fig:sampletraj_k4}
for low values of $k_3$ and $k_4$ are not bistable states but are
rather intrinsically unstable and transient states, each of which will
inevitably give rise to the other after a certain (stochastically
determined) period of time.  In this sense, our model is fundamentally
different from the bistable reaction networks which have previously
been discussed \cite{CA00,warren2004,dubnau2006}. On the other hand,
in the limit of rapid switch flipping, where $k_3$ or $k_4$ is large,
the mean-field description holds and the protein number distribution
does show a single peak whose position is well approximated by
Eq. (\ref{rho}), as shown in Figures \ref{fig:sampletraj_k3} and
\ref{fig:sampletraj_k4} for the case $k_3=1$.

\section{Steady--state statistics}

\subsection{Analytical solution}
Returning to the fully stochastic version of the reaction scheme
(\ref{eq:react}), we now present an exact solution for the
steady--state statistics of this model.  A solution for the case where
$\koff_3 = 0$ was sketched in Ref.~\cite{VAE08}. Here we present a
complete solution for the general case where $\koff_3 \ne 0$, and we
discuss the properties of the steady--state as a function of all the
parameters of the system.

We first define the probability $p_s(n,t)$ that the system has exactly
$n$ enzyme molecules at time $t$ and the switch is in the $s$ state
(where $s=\{\on,\off\}$). The time evolution of $p_s$ is described by
the following master equation:
\begin{multline}
\label{eq:master}
\frac{d p_s(n)}{d t}= (n+1) k_1 \ps(n+1)+ \ks_2 \ps(n-1) +n \kms_3
\pms(n)\\+ \kms_4 \pms(n) - (n k_1 + \ks_2 + n \ks_3 + \ks_4) \ps(n)
\,\,,
\end{multline}
where we use the shorthand notations $\{ \off,\on \} \equiv \{0,1\}$,
$\koff_2 \equiv 0$ and $\kon_2 \equiv k_2$.  In the steady state, the
time derivative in Eq.(\ref{eq:master}) vanishes, and the problem
reduces to a pair of coupled equations for $\pon$ and $\poff$:
\begin{subequations}
\label{eq:masterp}
\begin{multline}
\label{eq:masterpon}
(n+1) k_1 \pon(n+1)+ k_2 \pon(n-1) +n \koff_3 \poff(n)+ \koff_4
\poff(n) \\= (n k_1 + k_2 + n \kon_3 + \kon_4) \pon(n) \,\,,
\end{multline}
\vspace*{-0.7cm}
\begin{multline}
\label{eq:masterpoff}
 (n+1) k_1 \poff(n+1) + n \kon_3 \pon(n,t) + \kon_4 \pon(n,t) \\= (n
k_1 + n \koff_3 + \koff_4) \poff(n,t)\,\,\,.
\end{multline}
\end{subequations}
To solve the above equations we introduce the generating
functions
\begin{equation}
\label{eq:genfun}
G_s (z)= \sum_{n=0}^{\infty}  p_s(n) z^n\,\,.
\end{equation}
The steady-state equations (\ref{eq:masterp}) can be now written as a
set of linear coupled differential equations for $G_s$:
\begin{subequations}
\label{eq:gz}
\begin{gather}
\label{eq:gon}
\cL_1 \Gon(z) = \cL_2 \Goff(z)\,\,,\\
%\end{equation}
%\begin{equation}
\label{eq:goff}
\cL_3 \Goff(z) = \cL_4 \Gon(z)\,\,,
\end{gather}
\end{subequations}
where $\cL_i$ are linear differential operators:
\begin{subequations}
\begin{flalign}
\cL_1(z) =& k_1 (z-1) \p_z - k_2 (z-1) + \kon_3 z \p_z + \kon_4\,\,,\\
\cL_2(z)=&\koff_3 z \p_z + \koff_4 \,\,,\\
\cL_3(z)=& k_1 (z-1) \p_z + \koff_3 z \p_z + \koff_4\,\,,\\ 
\cL_4(z)=& \kon_3 z \p_z + \kon_4\,\,. 
\end{flalign}
\end{subequations}
In order to solve the two coupled \eqs~(\ref{eq:gz}) it is first
useful to take their difference. After simplification this yields
the relation:
\begin{equation}
\label{eq:relgon}
\p_z \Goff(z) = - \p_z \Gon(z) + \frac{k_2}{k_1} \Gon(z)\,\,.
\end{equation}
Next, we take the first derivative of
(\ref{eq:goff}) and then replace the derivatives of $\Goff$ with the
relation (\ref{eq:relgon}). After some algebra, one finds that $\Gon$
verifies the following second order differential equation:
\begin{multline}
\label{eq:diff}
k_1 (\alpha z - k_1 ) \Gon''(z) + (k_1 \beta - \gamma z )
\Gon'(z) - \delta \Gon(z)=0\,\,,
\end{multline}
where the Greek letters are combinations of the parameters of the
model:
\begin{subequations}
\begin{align}
\alpha &= k_1 + \kon_3 + \koff_3\,\,,\\
\beta & =k_1+k_2+\koff_3 + \kon_3 + \koff_4 + \kon_4\,\,,\\ 
\gamma &=k_2 (k_1 + \koff_3) \,\,,\\
\delta &= k_2 (k_1 + \koff_3 + \koff_4)\,\,.
\end{align}
\end{subequations}
We now introduce the new variable
\begin{equation}
u(z) \equiv u_z=\frac{\gamma}{k_1 \alpha} z - \frac{\gamma}{\alpha^2}
=u_0 + z (u_1-u_0)\,\,,
\label{udef}
\end{equation}
and the new parameter combinations:
\begin{equation}
\zeta=u_0 + \frac{\beta}{\alpha}\,\,, \qquad
\eta=\frac{\delta}{\gamma} \,\,.
\end{equation}
We can now write $\Gon(z)$ (and $\Goff(z)$) in terms of the variable
$u$ (\ref{udef}) by defining the functions
\begin{equation}
J_s(u)= G_s(z)\;.
\label{Jdef}
\end{equation}
The differential equation (\ref{eq:diff}) then reads:
\begin{equation}
u J_\on''(u) + (\zeta - u) J_\on'(u) - \eta J_\on(u) = 0\,\,.
\end{equation}
Looking for a regular power series solution of the form
\begin{equation}
J_\on(u)= \sum_{n=0}^{\infty} a_n u^n\,\,,
\label{Gonu}
\end{equation}
one obtains the following solution:
\begin{equation}
\label{eq:gonfull}
J_\on(u)=  a_0 \, _1F_1\left(\eta,\zeta, u \right)\,\,,
\end{equation}
where $_1F_1$ denotes the confluent hypergeometric function of the
first kind, 
\begin{equation}
_1F_1\left(\eta,\zeta, u \right) \equiv
\sum_{n=0}^{\infty} \frac{(\eta)_n}{(\zeta)_n}
\frac{u^n}{n !} 
\label{hgdef}
\end{equation}
and $(\alpha)_n=\alpha (\alpha+1) \dots (\alpha+n-1)$
denotes the Pochhammer symbol.

The constant $a_0$ will be determined using the boundary conditions,
which we discuss later. We first note that the above result for
$J_\on(u)$ can be translated into $G_\on(z)$ by replacing $u$ with the
expression of $u(z)$ in (\ref{Gonu}) and expanding in powers of $z$:
\begin{multline}
G_\on(z) = \sum_{n=0}^\infty a_n (u_0 + z (u_1-u_0))^n\\ 
= \sum_{n=0}^\infty a_n
\sum_{m=0}^{n} u_0^m [z (u_1-u_0)]^{n-m} \binom{n}{m}\\
= \sum_{n=0}^{\infty} z^n \sum_{m=n}^{\infty} a_m
u_0^{m-n} [(u_1-u_0)]^{n} \binom{m}{n}
\label{Gonsol}
\end{multline}
where we have relabelled the indices $n-m \to n$ and $n \to m$ in the
last line.  We can identify $\pon (n)$ from (\ref{eq:genfun}) as the
coefficient of $z^n$ in the above expression:
\begin{equation}
\label{eq:ponsol}
\pon(n)
=\sum_{m=n}^{\infty} a_m u_0^{m-n} (u_1 - u_0)^n \binom{m}{n}\;.
\end{equation}
From (\ref{Gonu}) and
(\ref{eq:gonfull}) we  read off
\begin{equation}
 a_n = \frac{a_0 }{n!}\ \frac{(\eta)_n}{(\zeta)_n}\, .
\label{an}
\end{equation}
Substituting (\ref{an}) in (\ref{eq:ponsol}) we deduce, using the
definition of the hypergeometric function (\ref{hgdef}) and noting
$(\alpha)_{n+m} = (\alpha)_n (\alpha+n)_m$, that
\begin{equation}
\pon(n)  =a_0 \frac{(u_1 - u_0)^n}{n!} \frac{(\eta)_n}{(\zeta)_n}
\,_1F_1(\eta+n, \zeta+n,u_0)\,\,.
\end{equation}
In deriving this expression we have, in fact, established the
following identity which will prove useful again later:
\begin{equation}
_1F_1(\eta, \zeta,u) = \sum_{n=0}^{\infty} \frac{z^n (u_1 -
u_0)^n}{n!} \frac{(\eta)_n}{(\zeta)_n} \,_1F_1(\eta+n,
\zeta+n,u_0)\,\,.
\label{Fident}
\end{equation}
To compute $\Goff(z)$, we integrate \eq (\ref{eq:relgon}), which
yields, using the form of $J_\on(u)$ (\ref{eq:gonfull}):
\begin{multline}
\label{eq:fastest}
\Goff(z) + \Gon(z) \\ - a_0 \frac{k_2 (\zeta-1)}{k_1 (\eta-1)
  (u_1-u_0)} \,_1F_1(\eta-1,\zeta-1,u_z)= \kappa\,\,,
\end{multline}
where $\kappa$ is our second integration constant. We then have two
constants, $a_0$ and $\kappa$, which still need to be determined. The
constant $\kappa$ can be found using the normalisation condition
$\sum_n (\pon(n)+\poff(n))=1$, which is equivalent to
$\Gon(1)+\Goff(1)=1$. Using this condition, we obtain
\begin{equation}\label{eq:kappa}
\kappa=1-a_0 \frac{k_2 (\zeta-1)}{k_1 (\eta-1) (u_1-u_0)}
 \,_1F_1(\eta-1,\zeta-1,u_1)\,\,.
\end{equation}
%and combining this with (\ref{eq:fastest}) an expression for
%$\Goff(z)$ in terms of $\Gon(z)$.  
In order to compute the remaining constant $a_0$, we consider the
boundary condition at $z=0$. From the definition (\ref{eq:genfun}) of
the generating function we see that $G_s(z=0) = p_s(n=0)$.  Our
boundary condition thus reads:
\begin{equation}
\label{eq:boundary}
J_\on(u_0)+J_\off(u_0)= \pon(0)+ \poff(0)\,\,.
\end{equation}
Setting $n=0$ in the master equation \eq (\ref{eq:masterpon}) [noting
that the term in $\pon(n-1)$ vanishes] gives $\poff(0)$ in terms of
$\pon(0)$ and $\pon(1)$:
\begin{equation}\label{eq:ponpoff}
\poff(0)=\frac{k_2+\kon_4}{\koff_4} \pon(0) - \frac{k_1}{\koff_4}
\pon(1)\,\,.
\end{equation}
Combining \eqs(\ref{eq:fastest}) [with $z=0$] and
(\ref{eq:kappa}), substituting in \eq (\ref{eq:boundary}), using \eq
(\ref{eq:ponpoff}) to eliminate $\poff(0)$, and finally substituting
in expressions for $\pon(0)$ and $\pon(1)$ from \eq (\ref{eq:ponsol}),
we  determine $a_0$:
\begin{multline}
\label{eq:a0sol}
a_0^{-1}=\left(1+ \frac{k_2+\kon_4}{\koff_4} \right) \,
_1F_1(\eta,\zeta,u_0) \\ - \frac{k_1 \eta (u_1-u_0)}{\koff_4 \zeta} \,
_1F_1(\eta+1,\zeta+1,u_0) \\ -\frac{k_2 (\zeta-1)}{k_1 (\eta-1)
(u_1-u_0)} \big[ \, _1F_1(\eta-1,\zeta-1,u_0) \\- \,
_1F_1(\eta-1,\zeta-1,u_1) \big] \,\,.
\end{multline}
The final step in obtaining our exact solution is to provide an
explicit expression for $\poff(n)$.  From (\ref{eq:fastest}) we have
\begin{multline}
\Goff(z) = \kappa - \Gon(z) \\ + a_0 \frac{k_2 (\zeta-1)}{k_1 (\eta-1)
  (u_1-u_0)} \,_1F_1(\eta-1,\zeta-1,u_z)\,\,,
\end{multline}
and using the identity (\ref{Fident})
we  obtain:
\begin{multline}
\label{eq:poffsol}
\poff(n)=\kappa \delta_{n,0} + \\ \frac{a_0}{n!} \bigg[\frac{k_2}{k_1}
(u_1-u_0)^{n-1} \frac{(\eta)_{n-1}}{(\zeta)_{n-1}} \, _1F_1(\eta+n-1,
\zeta+n-1,u_0)\\ - (u_1-u_0)^n \frac{(\eta)_n}{(\zeta)_n} \,_1F_1
(\eta+n,\zeta+n,u_0)\bigg]\,\,,
\end{multline}
where $\delta_{i,j}$ is the Kronecker delta. 

Our exact analytical
solution (\ref{eq:ponsol}), (\ref{eq:a0sol}) and (\ref{eq:poffsol}) is verified by comparison to computer simulation results in
the right panels of \figs \ref{fig:sampletraj_k3} and
\ref{fig:sampletraj_k4}. Here, we plot the probability distribution
function for the total number of enzyme molecules:
\begin{equation}
p(n)=\pon(n) + \poff(n)\;.
\label{p(n)}
\end{equation}
Computer simulations of the reaction set (\ref{eq:react}) were carried
out using Gillespie's stochastic simulation
algorithm~\cite{bortz75,gillespie1976}.  Perfect agreement is obtained
between the numerical and analytical solutions, as shown in \figs
\ref{fig:sampletraj_k3} and \ref{fig:sampletraj_k4}.

\subsection{Properties of the steady--state}\label{sec:pss}
Having derived the steady--state solution for $p(n)$, we now analyse
its properties as a function of the parameters of the model. We choose
to fix our units of time by setting $k_1$, the decay rate of enzyme
$R$, to be equal to unity (so our time units are $k_1^{-1}$). With
these units, the plateau value for the number of enzyme molecules in
the on switch state is given by $\rho_\on = k_2$. In this section, we
will only analyse the case where $\rho_\on = 100$. To further simplify
our analysis, we set $\kon_3=\koff_3=k_3$ and $\kon_4=\koff_4=k_4$ (a
discussion of the case where $\koff_3=0$ and $\kon_3 \neq 0$ is
provided in Ref. \cite{VAE08}).  We then analyse the probability
distribution $p(n)$ as a function of the $R$-dependent switching rate
$k_3$ and the $R$-independent switching rate $k_4$. The results are
shown in the right-hand panels of \fig \ref{fig:sampletraj_k3} and
\fig \ref{fig:sampletraj_k4}. We consider the three regimes discussed
in section \ref{sec:phen}: that in which enzyme number fluctuations
are much faster than switch flipping, that where the opposite is true,
and finally the regime where the two timescales are similar.

In the regime where switch flipping is much slower than enzyme
production/decay [$k_1 \gg \left(\kon_4 + k_2\kon_3/k_1\right)$], the
probability distribution $p(n)$ is bimodal.  This is easily
understandable in the context of the typical trajectories shown in the
left top panels in \figs \ref{fig:sampletraj_k3} and
\ref{fig:sampletraj_k4}: in this regime, the number of molecules of
$R$ always reaches its steady-state value before the next switch flip
occurs. It follows then that $\pon(n)$ is a bell-shaped distribution
peaked around $k_2/k_1$, while $\poff(n)$ is highly peaked around
zero, so that the total distribution $p(n)=\pon(n) + \poff(n)$ is
bimodal.

In contrast, when switching occurs much faster than enzyme number
fluctuations the probability distribution $p(n)$ is unimodal and bell
shaped, as might be expected from the trajectories in the bottom left
panels of \figs \ref{fig:sampletraj_k3} and
\ref{fig:sampletraj_k4}. As discussed in section \ref{sec:phen}, in
this regime the number of $R$ molecules behaves as a standard
birth-death process with effective birth rate given by $k_2$
multiplied by the average time the switch spends in the on state, and
death rate $k_1$.  For such a birth-death process the steady state
probability $p(n)$ is a Poisson distribution with mean given by the
ratio of the birth rate to the death rate. To show that our analytical
result reduces to this Poisson distribution, we consider the case
where enzyme-mediated switching dominates (as in \fig
\ref{fig:sampletraj_k3}), so that both $\koff_3$ and $\kon_3$ are much
greater than $k_1$. The fraction of time spent in the on state is
$\koff_3/\left(\kon_3+\koff_3\right)$, thus the effective birth rate
is $k_2 \koff_3/\left(\kon_3+\koff_3\right)$. In the limit $\kon_3 \to
\infty$ and $\koff_3 \to \infty$ with $r=\koff_3/\kon_3$ constant, one
finds that $\eta \to 1$, $\zeta \to 1$, and $u_z \to k_2 r z /[k_1
(1+r)]$. Using the fact that $\,_1F_1(1,1,x)=e^x$, \eq
(\ref{eq:gonfull}) gives, in this limit,
\begin{equation}
\label{eq:pppp}
\Gon(z)=a_0 \exp \left(\frac{k_2 r z }{k_1 (1+r)} \right)\,\,,
\end{equation}
which is the generating function of a Poisson distribution with mean
$k_2 \koff_3/[k_1 (\kon_3+\koff_3)]$. Plugging this result into \eq
(\ref{eq:fastest}) and taking again the limit $k_3 \to \infty$ [and
using that $\,_1F_1(0,0,x)=1$] finally yields the result that $p(n) =
\pon(n) + \poff(n)$ is indeed a Poisson distribution. The same
approach can be taken for the case of \fig \ref{fig:sampletraj_k4},
where $k_3$ is constant, and $\kon_4$ and $\koff_4$ become very
large. The probability distribution $p(n)$ then becomes a Poisson
distribution with mean $k_2 \koff_4/[k_1 (\kon_4+\koff_4)]$.  The
above result is only valid when $r \ne 0$. In fact, as shown in \fig
\ref{fig:sampletraj_k3off}, when $r=0$ the distribution of $R$ is
peaked at 0 and does not have a Poisson-like shape.

Finally, when there is no clear separation of timescales between
enzyme number fluctuations and switch flipping, the distribution
function for the number of enzyme molecules has a highly non-trivial
shape, as shown in the middle panels of \figs \ref{fig:sampletraj_k3}
and \ref{fig:sampletraj_k4}.

\section{First passage time distribution}
\label{sec:firstpassage}
We now calculate the first passage time distribution for our model
switch. We define this to be the distribution function for the amount
of time that the switch spends in the on or off states before
switching. This distribution is biologically relevant, since it may be
advantageous for a cell to spend enough time in the on state to
synthesise and assemble the components of the ``on'' phenotype (for
example, fimbriae), but not long enough to activate the host immune
system, which recognises these components. The calculation for the
case $\koff_3=0$ was sketched in \cite{VAE08}. Here we provide a
detailed calculation of the flip time distribution in the more general
case $\koff_3 \ne 0$. We find that this dramatically reduces the
parameter range over which the flip time distribution has a peak. We
demonstrate an important relation between the flip time distributions
for the two relevant choices of initial conditions (Switch Change
Ensemble and Steady State Ensemble). The first passage time
distribution is important and interesting from a statistical physics
point of view as it is related to ``persistence''. Generally,
persistence is expressed as the probability that the local value of a
fluctuating field does not change sign up to time $t$
\cite{Majumdar99}.  For the particular case of an Ising model,
persistence is the probability that a given spin does not flip up to
time $t$.  In our model, the switch state $S$ plays the role of the
Ising spin. For other problems, there has been much interest in the
long-time behaviour of the persistence probability, which can often
exhibit a power-law tail.  In our case, however, we expect an
exponential tail for the distribution of time spent in the on state,
because linear feedback will cause the switch to flip back to the off
state after some characteristic time. We are therefore interested not
only in the tail of the first passage time distribution, but in its
shape over the whole time range.

\subsection{Analytical results}
We consider the probability $F_{\s}(T|n_0) \dd T$ that if we begin
monitoring the switch at time $t_0$ when there are $n_0$ molecules of
the flipping enzyme $R$, it remains from time $t_0 \to t_0+T$ in state
$\s$, and subsequently flips in the time interval $t_0+T \to t_0+T+\dd
T$. This probability is averaged over a given ensemble of initial
conditions, determined by the experimental protocol for monitoring the
switch.  Mathematically, the initial condition $n_0$ for switch state
$\s$ is selected according to some probability $W_{\s}(n_0)$ and we
define
\begin{equation}
F_{\s}(T) = \sum_{n_0} F_{\s}(T|n_0) W_{\s}(n_0)
\end{equation}
as the flip time distribution for the ensemble of initial
conditions given by $W_{\s}(n_0)$.

The most obvious  protocol would be to
measure the interval $T$ from the moment of switch flipping, so that
the times $t_0$ correspond to switch flips and the $T$ are the
durations of the on or off switch states. We call this the {\em{Switch
Change Ensemble}} ($\sce$). In this ensemble, the probability
$W^\sce_s$ of having $n$ molecules of $R$ at the time $t_0$
when the switch flips into the $\s$ state is:
\begin{equation}\label{eq:w1}
W^\sce_s(n)=\frac{p_{1-s}(n) (n \kms_3 + \kms_4)}{\sum_n p_{1-s}(n) (n
\kms_3 + \kms_4)}\,\,.
\end{equation}
where for notational simplicity, $s=\{1,0\}$ represents
$\{{\mathrm{on,off}}\}$. The numerator of the r.h.s of
\eq(\ref{eq:w1}) gives the steady state probability that there are $n$
molecules present in state $1-s$, multiplied by the flip rate into
state $s$. The denominator normalises $W^\sce_s(n)$.

We also consider a second choice of initial condition, which we denote
the {\em Steady State Ensemble} ($\sse$). Here, the initial time $t_0$
is chosen at random for a cell that is in the $\s$ state. This choice
is motivated by practical considerations: experimentally, it is much
easier to pick a cell which is in the $\s$ state and to measure the
time until it flips out of the $\s$ state, than to measure the entire
length of time a single cell spends in the $\s$ state. The probability
$W^\sse_s$ of having $n$ molecules of $R$ at time $t_0$ is then
the (normalised) steady-state distribution for the $\s$ state:
\begin{equation}\label{eq:w2}
W^\sse_s=\frac{p_{s}(n)}{\sum_n p_{s}(n) }\,\,.
\end{equation}

To compute the distribution $F(T)$, we first consider the survival
probability $h_s^W(n,t)$, that, given that at time $t=0$ (chosen
according to ensemble $W$), the switch was in state $\s$, at time $t$
it is still in state $\s$ and has $n$ molecules of enzyme $R$.  As the
ensemble $W$ only enters through the initial condition, we may drop
the superscript $W$ in what follows. The evolution equation for $h_s$
is the same as for $p_s(n,t)$, but without the terms denoting switch
flipping into the $s$ state. This removes the coupling between $\pon$
and $\poff$ that was present in the evolution equations
(\ref{eq:masterp})):

\begin{subequations}
\label{eq:survn}
\begin{multline}
\label{eq:survonn}
\frac{\p}{\p t} \hon(n,t)  = (n+1) k_1 \hon(n+1,t) + k_2
\hon(n-1,t) \\- (n k_1 + k_2 + n \kon_3 + \kon_4) \hon(n,t) \,\,\,,
\end{multline}
\begin{multline}
\label{eq:survoffn}
\frac{\p}{\p t} \hoff(n,t) = (n+1) k_1 \hoff(n+1,t) - \\(n k_1 + n
\koff_3 + \koff_4)\hoff(n,t)\,\,\,.
\end{multline}
\end{subequations}
Introducing the generating function 
\begin{equation}
\htd_s(z,t)=\sum_{n=0}^\infty z^n h_s(n,t)\;,
\end{equation}
the above equations reduce to:
\begin{subequations}
\label{eq:surv}
\begin{multline}
\label{eq:survon}
\frac{\p}{\p t} \hont(z,t) =( k_1 - (k_1 + \kon_3  ) z) \p_z
\hont(z,t) \\+ (k_2 z - (k_2 + \kon_4)) \hont(z,t)  \,\,\,, 
\end{multline}
\begin{multline}
\label{eq:survoff}
\frac{\p}{\p t} \hofft(z,t) = ( k_1 - (k_1 + \koff_3 ) z) \p_z
\hofft(z,t) \\- \koff_4 \hofft(z,t)\,\,\,.
\end{multline}
\end{subequations}

We can relate $h$ to $F$ by noting that $\sum_n h_s(n,t) =
\htd_s(1,t)$ is the total probability that the switch has not flipped
up to time $t$.  Hence,
\begin{equation}
F_s(t)=- \p_t \htd_s (1,t)\;.
\label{eq:link}
\end{equation}
Equations (\ref{eq:surv}) can be solved using the method of
characteristics~\cite{courant}. The result, detailed in
Appendix~\ref{app:char}, is: 
\begin{multline}
\label{eq:hton}
\hont(z,t)= e^{- \omega_\on t} e^{k_2 \tau_{\on}
(z-k_1 \tau_{\on}) (1- e^{-t/\tau_{\on}})}\\ \times \tW(k_1 \tau_{\on}
+ e^{-t/\tau_{\on}}(z-k_1 \tau_\on)) \,\,\,,
\end{multline}
where $\tau_\on = (k_1+\kon_3)^{-1}$ and $\omega_\on=\kon_4+ k_2 (1-
k_1 \tau_{\on})$. The function $\tW$ is the generating function for
the distribution of enzyme numbers $W(n)$ at the starting time for the
measurement:
\begin{equation}\label{eq:ww}
\tW(z) = \sum_n W(n) z^n\,\,,
\end{equation}
where $W$ refers to $W^\sce$ or $W^\sse$.  The function $\hofft(z,t)$
can be obtained in an analogous way: this produces the same expression
as for $\hont$, but with $k_2$ set to zero and with all ``$\on$''
superscripts replaced by ``$\off$'':
\begin{equation}
\label{eq:htoff}
\hofft(z,t)= e^{- \koff_4 t} \tW(k_1 \tau_{\off} +
 e^{-t/\tau_{\off}}(z-k_1 \tau_\off)) \,\,,
\end{equation}
so that $\tau_\off = (k_1+\koff_3)^{-1}$. We can then obtain
the distributions $F_{\on}(T)$ and $F_\off(T)$ by differentiating the
above expressions, according to Eq.(\ref{eq:link}):
\begin{multline}
\label{eq:font}
F_{\on}(T)=\exp \left(-(\omega_{\on}+1/\tau_{\on}) T +k_2 \tau_\on
(1-e^{-T/\tau_\on}) \right)\\ \times \Bigg\{ \Big[\omega_\on
e^{T/\tau_\on} + k_2 (k_1 \tau_\on -1) \Big] \tW \left( k_1 \tau_\on +
e^{-T/\tau_\on} (1-k_1 \tau_\on) \right)\\ +\left(\frac{1}{\tau_\on} -
k_1 \right) \tW' \left( k_1 \tau_\on + e^{-T/\tau_\on} (1-k_1
\tau_\on) \right) \Bigg\}\,\,,
\end{multline}
\begin{multline}
\label{eq:fofft}
F_\off(T)= \exp \left(-(\koff_4 + 1/\tau_\off) T \right)\\ \times
\Bigg\{ \koff_4 e^{T/\tau_\off} \tW \left(k_1 \tau_\off +
e^{-T/\tau_\off} (1-k_1 \tau_\off) \right)\\
+ \left(\frac{1}{\tau_\off} - k_1 \right) \tW' \left(k_1 \tau_\off +
e^{-T/\tau_\off} (1-k_1 \tau_\off) \right) \Bigg\}\,\,.
\end{multline}

In the above expressions, the function $\tW_s$ 
is given for  the steady state ensemble ($\sse$) by
\begin{equation}
{\tW_s}^\sse=G_s(z)/G_s(1)
\end{equation}
and for  the switch change ensemble ($\sce$) by
\begin{equation}
{\tW_s}^\sce(z)=\frac{\kms_3 z G'_{1-s}(z) + \kms_4
G_{1-s}(z)}{\kms_3 G'_{1-s}(1) + \kms_4 G_{1-s}(1)}\,.
\end{equation}

\subsection{Relation between SSE and SCE}

\label{sec:relation}
\begin{figure}
\includegraphics[width=\columnwidth]{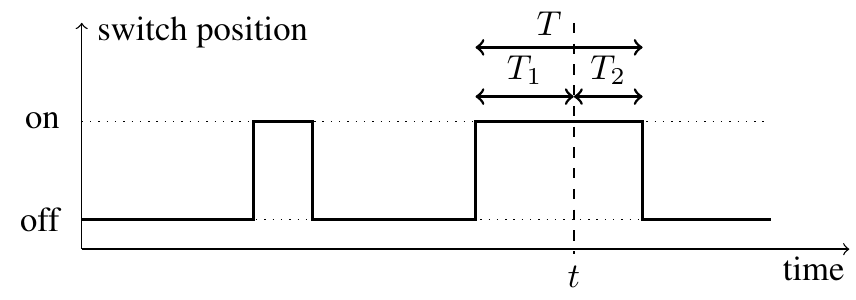}
\caption{
\label{fig:illustration} Schematic illustration of a possible
 time trajectory for the switch; $t$ is a random time falling
in an interval of total length $T$ and splitting it into two other
intervals denoted $T_1$ and $T_2$, as discussed in Section \ref{sec:relation}.}
\end{figure}

We now show that a useful and simple relation can be derived between
$F_\sse(T)$ and $F_{\sce}(T)$. Let us imagine that we pick a random
time $t$, chosen uniformly from the total time that the system spends
in state $\s$. The time $t$ will fall into an interval of duration
$T$, as illustrated in \fig \ref{fig:illustration}. We can then split
the interval $T$ into the time $T_1$ before $t$ and the time $T_2$
after $t$, such that $T_1+T_2=T$.

We first note that the probability that our randomly chosen time $t$
falls into an interval of length $T$ is:
\begin{equation}
\label{eq:plength}
\Prob(T)\, dT= \frac{T \,F_s^\sce(T) \, dT}{\int_0^\infty T'
\,F_s^\sce(T')\,dT' } 
%= \frac{T \,F^\sce(T)_s \,dT}{\avg{T}_\sce}\,dT \,\,.
\end{equation}
Eq.(\ref{eq:plength}) expresses the fact that the probability
distribution for a randomly chosen flip time $T$ is $F^\sce_s(T)
\,dT$, but the probability that our random time $t$ falls into a given
segment is proportional to the length of that segment. Since the time
$T$ is chosen uniformly, the probability distribution for $T_2$, for a
given $T$, will also be uniform (but must be less than $T$):
\begin{equation}\label{eq:cond}
\Prob(T_2|T)\,\,dT=\frac{\Theta(T-T_2)}{T} \,\,dT\,\,.
\end{equation}
One can now obtain $F^\sse_s$ from $\Prob(T_2|T)$ by integrating
Eq.(\ref{eq:cond}) over all possible values of $T$, weighted by the
relation (\ref{eq:plength}). This leads to the following relation
between $F^\sce$ and $F^\sse$:
\begin{equation}
\label{eq:linkbis}
F^\sse_s(T_2)=\frac{\int_{T_2}^\infty F^\sce_s(T')
\,\,dT'}{\int_0^\infty T' F^\sce_s(T') \,\,dT'}\,\,.
\end{equation}
Taking the derivative with respect to $T_2$ this can be recast as 
\begin{equation}
\label{eq:linkbis2}
\frac{d F^\sse_s(T)}{d T}=-\frac{F^\sce_s(T)}{
\langle T \rangle_{\rm SCE}}
\end{equation}
where $\langle T \rangle_{\rm SCE}$ is simply the mean duration of a
period in the on state. We have verified numerically that the
expressions (\ref{eq:font}) and (\ref{eq:fofft}) for $F^\sse_s(T)$ and
$F^\sce_s(T)$ derived above do indeed obey the relation
(\ref{eq:linkbis2}).  This relation can also be understood in terms of
backward evolution equations as we discuss in Appendix \ref{sec:bev}.

\subsection{Presence of a peak in $F(T)$}
We now focus on the shape of the flip time distribution $F(T)$, in
particular, whether it has a peak.  A peak in $F^{\sce}_{\on}(T)$
could be biologically advantageous for two complementary reasons.
Firstly, after the switch enters the on state there may be some
start-up period before the phenotypic characteristics of the on state
are established, so it would be wasteful for flipping to occur before
the on state of the switch has become effective.  Secondly, the on
state of the switch may elicit a negative environmental response, such
as activation of the host immune system, so that it might be
advantageous to avoid spending too long a time in the on state. For
example, in the case of the {\em{fim}} switch, a certain amount of
time and energy is required to synthesise fimbriae, and this effort
will be wasted if the switch flips back into the off state before
fimbrial synthesis is complete.  On the other hand, too large a
population of fimbriated cells would trigger an immune response from
the host, therefore the length of time each cell is in the fimbriated
state needs to be tightly controlled. We note that for bistable
genetic switches and many other rare event processes, waiting time
distributions are exponential (on a suitably coarse-grained
timescale). This arises from the fact that the alternative stable
states are time invariant in such systems. The presence of a peak in
$F^{\sce}_{\on}(T)$ for our model switch would indicate fundamentally
different behaviour, which occurs because the two switch states in our
model are time-dependent.

The presence of a peak in the distribution $F(T)$ requires the slope
of $F(T)$ at the origin to be positive. Applying this condition to the
function $F_\on$ (\ref{eq:font}) we get: 
\begin{multline}
\label{eq:inequality1}
(k_2 \kon_3 - (\kon_4)^2) \tW(1) - \kon_3 (k_1 + \kon_3 + 2 \kon_4)
\tW'(1) \\- (\kon_3)^2 \tW''(1) >0\,\,.
\end{multline}
\eq (\ref{eq:ww}) allows us to expressing the derivatives of $\tW(1)$
as functions of the moments of $n$, so that we finally get our
condition as a relation between the mean and the variance of the
initial ensemble:
\begin{multline}
\label{eq:inequality}
 k_2 \kon_3 - (\kon_4)^2 - \kon_3 (k_1  + 2  \kon_4)
\avg{n}_{W_{\on}} \\- (\kon_3)^2 \avg{n^2}_{W_\on} >0\,\,,
\end{multline}
where $\avg{\dots}_{W_\on}$ denotes an average taken using the weight
$W_\on$ of Eq. (\ref{eq:w1}) or (\ref{eq:w2}). Analogous conditions
can be found for a peak in the off to on waiting time
distribution. The moments involved in the above inequality can be
computed using the exact results of the previous section. The
l.h.s. of (\ref{eq:inequality}) can then be computed numerically for
different values of the parameters, to determine whether or not a peak
is present in $F(T)$.

For the SSE, there is never a peak in the flip time distribution. This
follows directly from the relation (\ref{eq:linkbis2}) between the SSE
and SCE, which shows that the slope of $F^\sse_s(T)$ at the origin is
always negative:
\begin{equation}
\left. \frac{d F^\sse_s(T)}{d T}\right|_{T=0} = -
\frac{F^\sce_s(0)}{\avg{T}_\sce} <0\,\,.
\end{equation}
Thus a peak in the waiting time distribution cannot occur when the
initial condition is sampled in the steady state ensemble.

For the SCE, we tested inequality (\ref{eq:inequality}) numerically
and found that a peak in the distribution $F(T)$ is possible for the
time spent in the on state ($F_\on^\sce$), but not for the off to on
waiting time distribution ($F_\off^\sce$). This is as expected and can
be explained by noting that to produce a peak in $F_s^\sce(T)$, the
flipping rate must increase with time in state $s$. In the on state
the flipping rate typically does increase with time as the enzyme $R$
is produced, while in the off state the flipping rate decreases in
time as $R$ decays.

\begin{comment}
An intuitive explanation can be provided for the
peak in $F_\on^\sce$. Once the switch enters the on state, the amount
of flipping enzyme $R_1$ begins to increase, and the on to off
flipping rate increases concomitantly. The probability of flipping
shortly after entering the on state is thus lower than the probability
of flipping later, once more molecules of $R_1$ are present. Depending
on the parameter values, this may lead to a peak in the distribution
$F_\on^\sce$.

%In contrast, when the initial condition is sampled in the SSE, the
%flipping rate is typically already high, since a significant amount of
%$R_1$ is already present. We therefore expect that one is less
%likely to find a peak in $F_\on^\sse$ than in $F_\on^\sce$; however we
%are unable to explain intuitively why it is in fact impossible to find
%a peak in $F_\on^\sse$.

The fact that no peak is present in either the SCE ensemble for the
off to on flip time distribution $F_\off(T)$ might be due to the fact
that when the switch is in the off state, the amount of flipping
enzyme $R_1$ and hence the flipping rate {\em decreases} after the on
to off flip. Therefore, the probability of flipping shortly after the
previous flip is higher than the probability of flipping later, when
the amount of $R_1$ has dropped off, and no peak is expected.
\end{comment}

We now discuss the general conditions for the occurrence of a peak in
$F_\on^\sce$. We first recall from section \ref{sec:pss} that in the
regime where the copy number of the enzyme $R$ relaxes much faster
than the switch flips [$k_1 \gg \kon_4 + k_2\kon_3/k_1$], the plateau
level of $R$ is reached rapidly after entering the on state, so that
the flipping rate out of the on state is essentially constant. This
leads to effectively exponentially distributed flip times from the on
state, so that no peak is expected. In the opposite regime, where
switch flipping is much faster than $R$ number relaxation [$k_3 \gg
0$], we again expect Poissonian statistics and therefore exponentially
distributed flip times.  Thus it will be in the intermediate range of
$k_3$ that a peak in the flip time distribution may occur.  The exact
condition for this (\ref{eq:inequality}) is not particularly
transparent as the dependence on the parameters is implicit in the
values of the $\avg{n}_{W_{\on}}$ and $\avg{n^2}_{W_{\on}}$.  In
particular, the effects of the parameters $k_3$ and $k_2$ are coupled,
since the effective $R$-mediated switching rate depends on the copy
number of $R$. However we can make a broadbrush description of what is
required. First the switch should enter the on state with typical
values of $n \ll \rho_{\rm on}$ so that there is an initial rise in
the value of $n$ and therefore the flipping rate. Second, we expect
that the flipping should be predominantly effected by the enzyme $R$
rather than spontaneously flipping {\em{i.e.}}  $k_3$ should govern
the flipping rather than $k_4$.

\begin{figure}
\includegraphics[width=\columnwidth,clip=true]{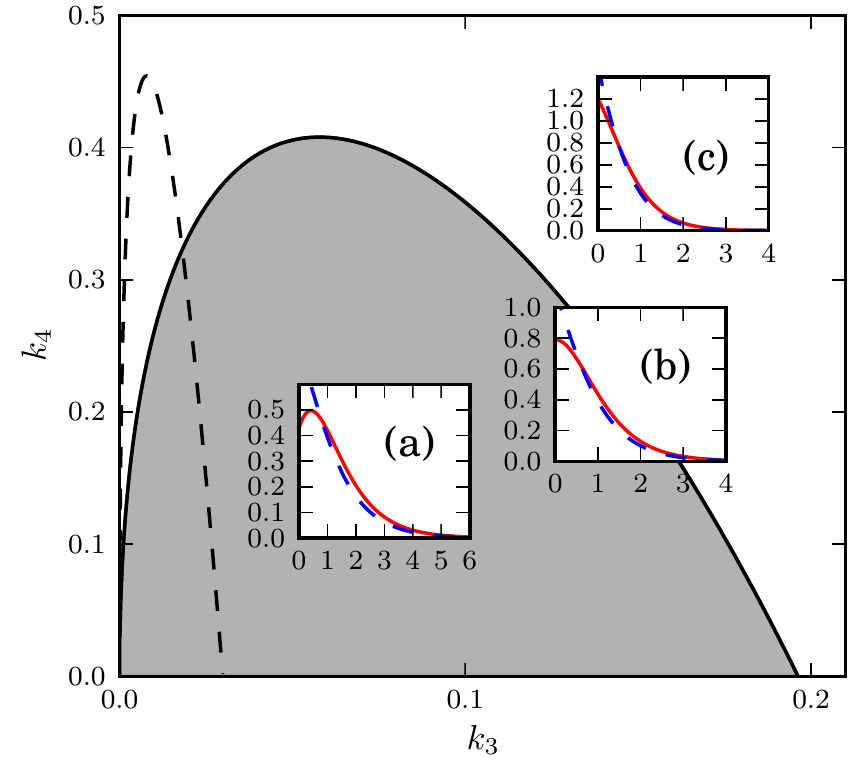}
\caption{\label{fig:diagram} Occurrence of a peak in the waiting time
distribution sampled in the Switch Change Ensemble. The shaded area
delimits the region where there is a peak (here the parameters are:
$k_1 =1$, $k_2=10$ and $\koff_3=\kon_3=k_3$ and $\koff_4=\kon_4=k_4$).
The dashed line delimits the same region for $k_2=100$. The insets
show an instance of the distribution both in the SCE (solid red line)
and in the SSE (blue dashed line): (a) There is a peak ($k_2=10$,
$k_3=0.1$, $k_4=0.1$); (b) On the transition line, where the slope at
the origin vanishes ($k_2=10$,$k_3=0.15$, $k_4=0.209384...$); (c)
There is no peak ($k_2=10$, $k_3=0.2$, $k_4=0.35$).}
\end{figure}

\fig \ref{fig:diagram} shows the region in the $k_3$--$k_4$ plane
where $F_\on^\sce$ has a peak, for the case where $\kon_3 = \koff_3 =
k_3$ and $\kon_4 = \koff_4 = k_4$. These results are obtained
numerically, using the inequality (\ref{eq:inequality}). The
distribution $F_\on^\sce$ is peaked for parameter values inside the
shaded region. The insets show examples of the distributions
$F_\on^\sce(T)$ and $F_\on^\sse(T)$ for various parameter values. At
the boundary in parameter space between peaked and monotonic
distributions (solid line in \fig \ref{fig:diagram}), $F_\on^\sce(T)$
has zero gradient at $T=0$ (inset (b)). The dashed line in \fig
\ref{fig:diagram}) shows the position of the boundary for a larger
value of the enzyme production rate $k_2$. As $k_2$ increases, the
range of values of $k_3$ for which there is a peak
decreases. Increasing $k_2$ increases the number of enzyme present,
which will increase both the off to on and on to off switching
frequency, since here $\kon_3=\koff_3=k_3$. Thus it appears that
approximately the same qualitative behaviour can be obtained for
smaller values of $k_3$ when $k_2$ is increased.

\begin{figure}
\includegraphics[width=\columnwidth,clip=true]{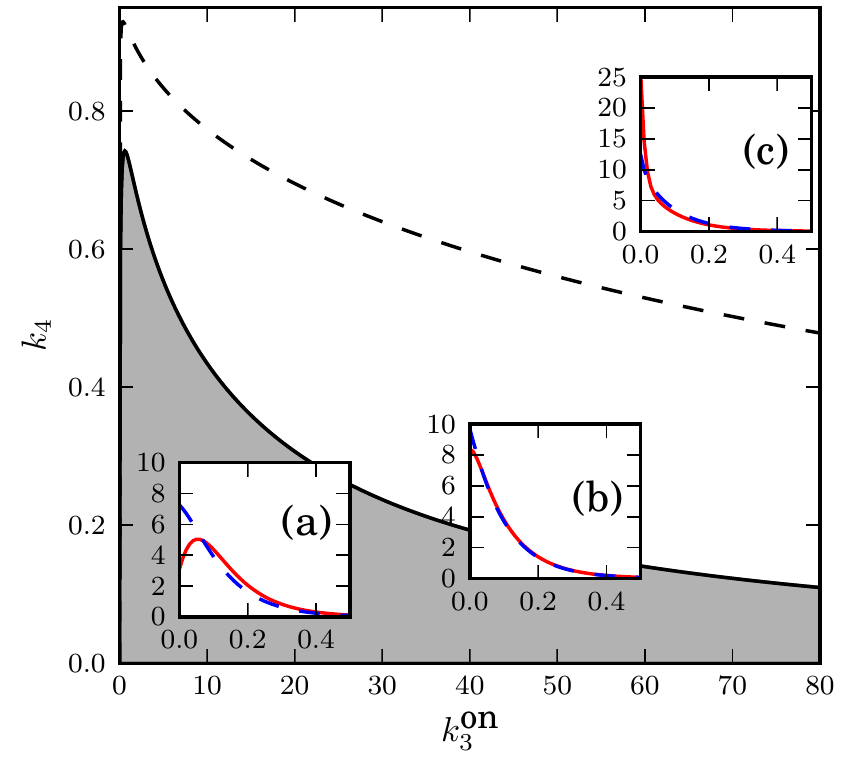}
\caption{\label{fig:diagramk3off0} Same plot as \fig \ref{fig:diagram}
but for $\koff_3=0$. The shaded area delimits the values of $k_4$ and
$\kon_3$ (with $k_2=10$) for which there is a peak in the flip time
distribution. The dashed line is the separation line for
$k_2=100$. The examples in the insets have as parameters $k_2=10$ and:
(a) $\kon_3=15$, $k_4=0.15$; (b) $\kon_3=50$, $k_4=0.162383...$; (c)
$\kon_3=80$, $k_4=0.4$.  }
\end{figure}

In our previous paper \cite{VAE08}, we analysed the case where
$\koff_3=0$: {\em{i.e.}} the flipping enzyme $R$ switches only in the
on to off direction. This case applies to the {\em{fim}} system. \fig
\ref{fig:diagramk3off0} shows the analogous plot, as a function of
$\kon_3$ and $k_4$, when $\koff_3=0$. The region of parameter space
where a peak occurs in $F_\on^\sce(T)$ is much wider than for nonzero
$\koff_3$. In this case an increase of $k_2$ produces a {\em{larger}}
range of parameter values $\kon_3$ for which there is a peak (dotted
line in \fig \ref{fig:diagramk3off0}). Here, the off to on
switching process is $R$-independent, and is mediated by $k_4$ only
(since $\koff_3=0$). The typical initial amount of $R$ present on
entering the on state is thus not much affected by $k_2$, although the
plateau level of $R$ increases with $k_2$. Therefore, as $k_2$
increases, the enzyme copy number in the on state becomes more
time-dependent, increasing the likelihood of finding a peak.

\begin{figure}
\includegraphics[width=\columnwidth,clip=true]{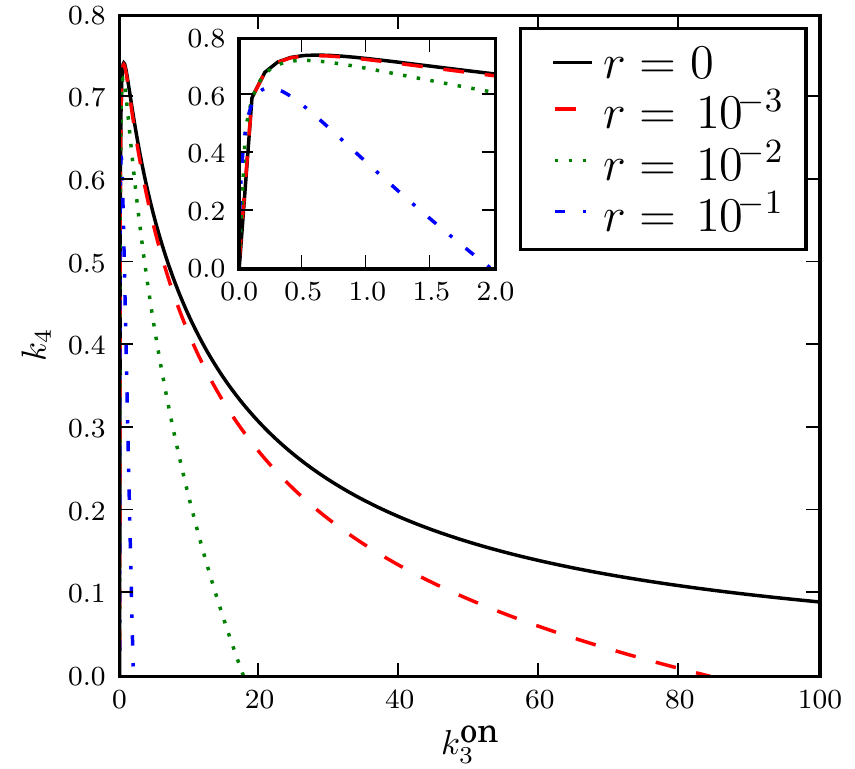}
\caption{\label{fig:diagramk_vary_3off} Diagram showing the occurrence
of a peak when the ratio $r=\koff_3/\kon_3$ is varied. Here $k_1=1$
and $k_2=10$. The inset shows a zoom of the plot in the vicinity of
$\kon_3=0$.}
\end{figure}

The comparison between \figs \ref{fig:diagram} and
\ref{fig:diagramk3off0} suggests that the relative magnitudes of the
$R$-mediated switching rates in the on to off and off to on
directions, $\kon_3$ and $\koff_3$, play a major role in determining
the parameter range over which $F_\on^\sce$ is peaked. This
observation is confirmed in \fig \ref{fig:diagramk_vary_3off}, where
the boundary between peaked and unpeaked distributions is plotted in
the $\kon_3$--$k_4$ plane for various ratios $r=\koff_3/\kon_3$. The
larger the ratio $r$, the smaller the region in parameter space where
there is a peak. An intuitive explanation for this might be that as
$r$ increases, the the typical initial number of $R$ molecules in the
on state increases, so that less time is needed for the $R$ level to
reach a steady state, resulting in a weaker time-dependence of the on
to off flipping rate and less likelihood of a peak occurring in
$F(T)$. If the presence of a peak in $F_\on^\sce$ is indeed an
important requirement for such a switch in a biological context, then
we would expect that a low value of $\koff_3$, as is in fact observed
for the {\em{fim}} system, would be advantageous.

\section{Correlations}

A peaked distribution of waiting times is by no means the only
potentially useful characteristic of this type of switch. In this
section, we investigate two other types of behaviour that may have
important biological consequences: correlations between successive
flips of a single switch, and correlated flips of multiple switches in
the same cell. We analyse these novel phenomena using numerical
methods. We introduce a new correlation measure which enables us to
quantify the extent of the correlation as a function of the parameter
space. Our main findings are that a single switch shows time
correlations which appear to decay exponentially, and that two
switches in the same cell can show correlated or anti--correlated
flipping behaviour depending on the values of $\koff_3$ and $\kon_3$.

\subsection{Correlated flips for a single switch}
Biological cells often experience sequences of environmental changes:
for example, as a bacterium passes through the human digestive system
it will experience a series of changes in acidity and temperature. It
is easy to imagine that evolution might select for gene regulatory
networks with the potential to ``remember'' sequences of events. The
simple model switch presented here can perform this task, in a very
primitive way, because it produces correlated sequences of switch
flips: the amount of $R$ enzyme present at the start of a particular
period in state $\s$ depends on the recent history of the
system. 
In contrast, for bistable gene
regulatory networks, or other bistable systems,
successive flipping events are uncorrelated, as long as the
system has enough time to relax to its steady state between
flips.

 In our recent work \cite{VAE08}, we demonstrated that successive
switch flips can be correlated for our model switch, and that this
correlation depends on the parameter $\koff_3$: correlation increases
as $\koff_3$ increases. Here, we extend our study and introduce a new
measure of these correlations: the two time probability $p(s,t;
s',t')$ that the switch is in position $s$ at time $t$ and in position
$s'$ at time $t'$. In the steady state the two-time probability
depends only on the time difference $\tau=t-t'$. In order to compare
different simulations results, we define the auto-correlation
function:
\begin{equation}
C(\tau)=
\frac{p_{\on-\on}(\tau)}{p_\on}+\frac{p_{\off-\off}(\tau)}{p_\off} -1,
\label{Ctau}
\end{equation}
where $p_{\on-\on}(\tau) = p(\on,t; \on, t+\tau)$,
$p_{\off-\off}(\tau) = p(\off,t; \off, t+\tau)$, and $p_\on$
($p_\off$) is the probability of being in the $\on$ (off) state. The
correlation function (\ref{Ctau}) takes values between $-1$ and $1$,
in such a way that it is positive for positive correlations, negative
for negative correlations and vanishes if the system is uncorrelated.
This function allows us to understand whether, given that the switch
is in a given position $s$ at time $t$, it will be in the same state
$s$ at a later time $t+\tau$.

\begin{figure}
\includegraphics[width=\columnwidth,clip=true]{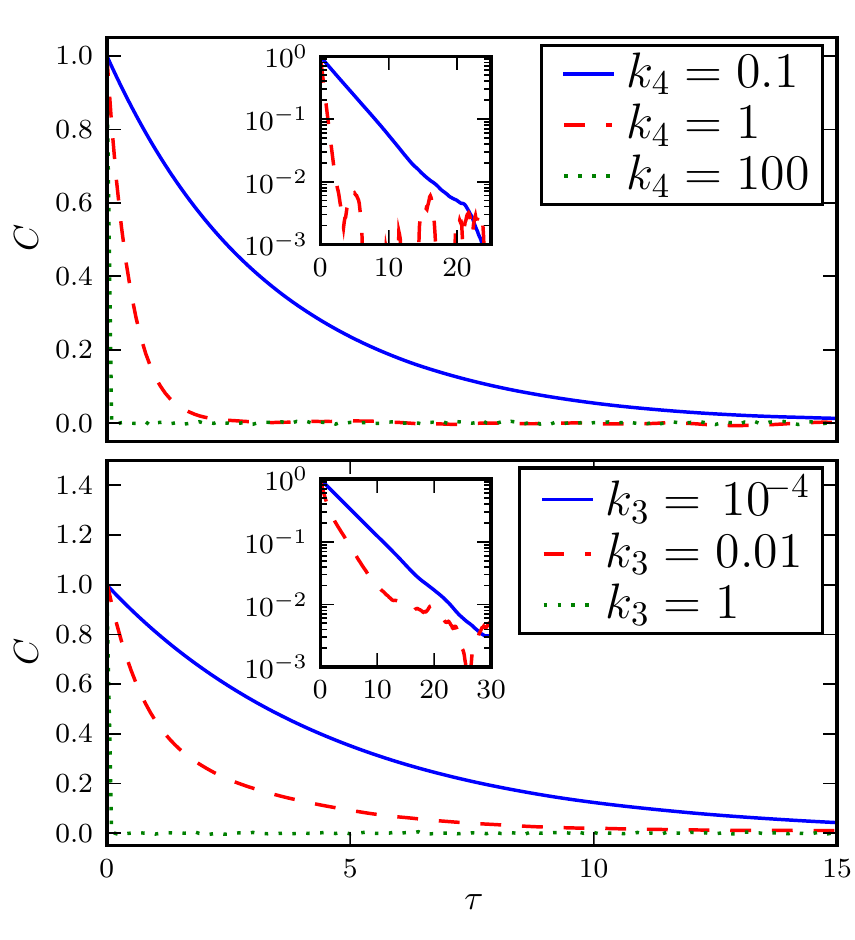}
\caption{ \label{fig:corrone}(colour online) The two-time
  auto-correlation function $C(\tau)$ for $k_1=1$, $k_2=100$. The
  insets shows the same data in a semi-log scale. \textsc{Top}: $k_4$
  is varied with constant $k_3=0.001$. \textsc{Bottom}: $k_3$ is
  varied with constant $k_4=0.1$. }
\end{figure}

\begin{figure}
\includegraphics[width=\columnwidth,clip=true]{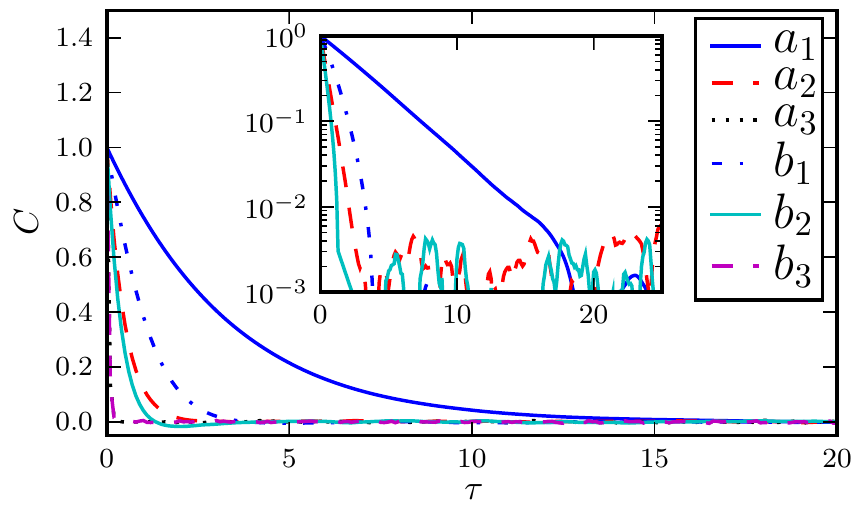}
\caption{ \label{fig:corronek3off}(colour online) The
  correlation function $C(\tau)$ when $\koff_3=0$. As
  previously, $k_1=1$ and $k_2=100$. The data labelled as $a$
  corresponds to $\kon_3=0.001$ while $b$ corresponds to
  $\kon_3=0.01$. For each $a$ and $b$ the superscripts $1$, $2$ and
  $3$ refer to different values of $k_4=0.1$, $1$ and $10$
  respectively. The inset shows the same plot on a semi-log scale.}
\end{figure}

\fig\ref{fig:corrone} shows simulation results for different values of
$\kon_3=\koff_3=k_3$ and $\kon_4=\koff_4=k_4$. As expected, the
correlation function vanishes in the limit of large $\tau$, meaning
that in this limit there are no correlations. Furthermore, we can see
that the strength of the correlations decreases when either $k_3$ or
$k_4$ are increased. This is consistent with the previous remark that
in the limit of large switching rate ({\em i.e.}  either $k_3$ or
$k_4$) the distribution of enzyme numbers tends to a Poisson
distribution. It is thus not surprising that in this same limit the
correlations vanish. In the insets of \fig\ref{fig:corrone} we plot
the same correlation function on a semi-logarithmic scale. The data
for the highest values of $k_3$ or $k_4$ (the dotted green curves) is
not shown since the decrease is too sharp, and does not allow for a
clear interpretation. For the smallest values of $k_3$ and $k_4$ (blue
curves), the decay seems to be exponential. However, for intermediate
values of $k_3$ or $k_4$ (dashed red curves) the evidence for an
exponential decay is less clear and the issue deserves a more
extensive numerical investigation. For the sake of completeness we
also show in figure \ref{fig:corronek3off} similar data for the case
where $\koff_3=0$. We find that qualitatively the data has a very
similar behaviour to the case where $\koff_3=\kon_3$.

\subsection{Multiple coupled switches}
Many bacterial genomes contain multiple phase-varying genetic
switches, which may demonstrate correlated flipping. For example, in
uropathogenic {\em{E. coli}}, the {\em{fim}} and {\em{pap}} switches,
which control the production of different types of fimbriae, have been
shown to be coupled \cite{holden2004,holden2006}. Although these two
switches operate by different mechanisms, it is also likely that
multiple copies of the same switch are often present in a single
cell. This may be a consequence of DNA replication before cell
division (in fast-growing {\em{E. coli}} cells, division may proceed
faster than DNA replication, resulting in up to $\sim 8$ copies per
cell). Randomly occurring gene duplication events, which are believed
to be an important evolutionary mechanism, might also result in
multiple copies of a given switch on the chromosome. It is therefore
important to understand how multiple copies of the same switch would
be likely to affect each other's function~\cite{ribeiro}.

Let us suppose that there are two copies of our model switch in the
same cell. Each copy contributes to and is influenced by a common pool
of molecules of enzyme $R$.  Our model is still described by the set
of reactions (\ref{eq:react}), but now the copy number of $S_\on$ and
$S_\off$ can vary between 0 and 2 (with the constraint that the total
number of switches is 2).

\begin{figure}
\includegraphics[width=\columnwidth,clip=true]{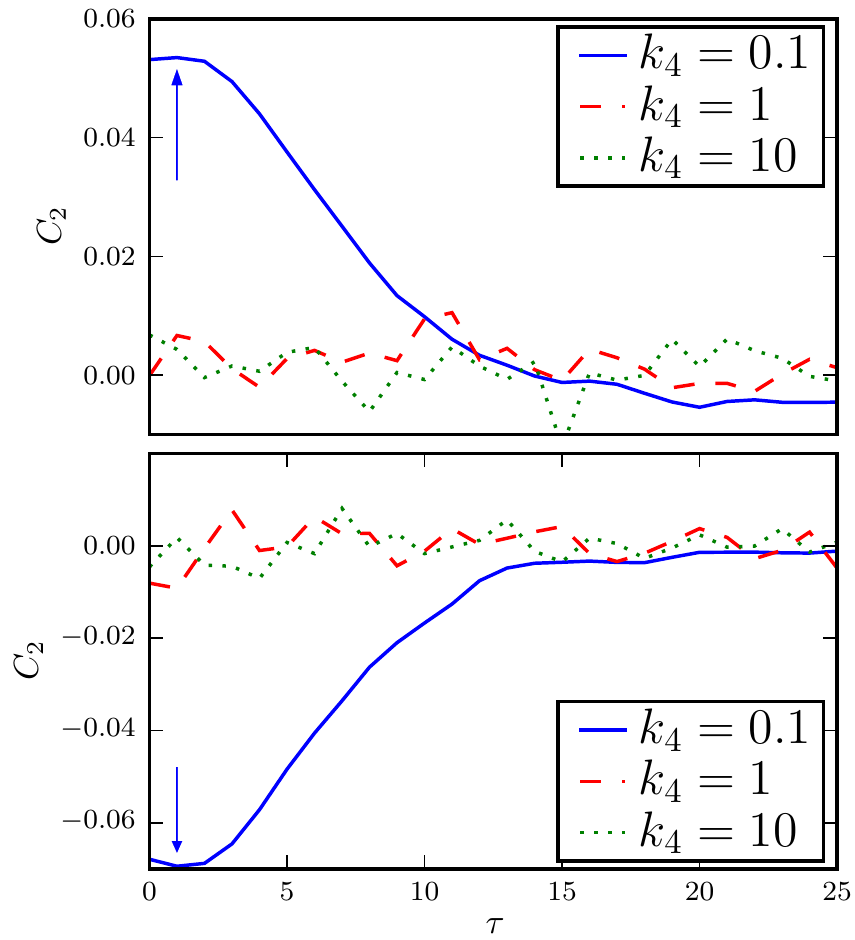}
\caption{\label{fig:twoswitches}(colour online) Normalised two-time
correlation function $C_2(\tau)$ for two identical switches. The
parameter values are: $k_1=1$, $k_2=100$,$\kon_3=0.001$. In the top
panel $\koff_3=\kon_3$ while in the bottom panel $\koff_3=0$. The
parameter $k_4$ is varied from $0.1$ to $100$ in each case.}
\end{figure}

To measure correlations between the states of the two switches
(denoted $s_1$ and $s_2$) we define the {\em two switch} joint
probability $p_2(s_1,t;s_2,t')$ as the probability that switch
1 is in state $s_1$ at time $t$ and switch $2$ is in
state $s_2$ at time $t'$. This function is the natural extension of
the previously defined two-time probability for a single switch. Thus,
in analogy to (\ref{Ctau}), we can define a two-time correlation
function:
\begin{equation}
C_2(\tau)= \frac{p_2(\on,t;\on,t+\tau)}{p_\on}
+\frac{p_2(\off,t;\off,t+\tau)}{p_\off}-1\,\,,
\end{equation}
where $p_\on$ ($p_\off$) is again the steady--state probability for a
single switch to be $\on$ ($\off$). If the two switches are completely
uncorrelated, we expect that $p_2(\on,t;\on,t') = p_\on^2$ and
$p_2(\off,t;\off,t') = p_\off^2$, so that $C_2(\tau) = 0$ (given that
$p_\on + p_\off=1$). In contrast, if the switches are completely
correlated, $p_2(\on,t;\on,t') = p_\on$, $p_2(\off,t;\off,t') =
p_\off$ and $C_2(\tau) = 1$. For completely anti-correlated switches,
we expect that $p_2(\on,t;\on,t') = p_2(\off,t;\off,t') = 0$, and
$C_2(\tau) = -1$. In \fig \ref{fig:twoswitches} we plot the function
$C_2(\tau)$ for two identical coupled switches, for several parameter
sets.  Our results show that for small values of $k_4$, there is
correlation between the two switches, over a time period $\approx 10
k_1^{-1}$, which is of the same order as the typical time spent in the
on state for these parameter values.  Our results also show that the
nature of these correlations depends strongly on $\koff_3$.  In the
case where $\koff_3=\kon_3$ (top panel of \fig \ref{fig:twoswitches}),
one can see that the correlation is positive, meaning that the two
switches are more likely to be in the same state. In contrast, when
$\koff_3$ is set to zero (bottom panel of \fig \ref{fig:twoswitches}),
the correlation is negative, meaning that the two switches are more
likely to be in different states.

To understand these correlations, consider the extreme situation where
both the two switches are off, and the number molecules of $R$ has
dropped to zero. In this case, the only possible event is a $k_4$
mediated switching which could take place, for instance, for the first
switch. Then, once the first switch is on, it will start producing
more enzyme, and, if $\koff_3 \ne 0$, this will enhance the
probability for the second switch to flip on too. This might explain
why, when $\koff_3=\kon_3$ we see a positive correlation between the
two switches. On the other hand, if we consider the opposite situation
where both the two switches are on, and the number of molecules of $R$
is around its plateau value, then the on to off switching probability
for the two switches will be at its maximum. However, after one of the
switches has flipped (e.g. the first), the switching probability will
start decreasing, this reducing the flipping rate for the second
switch. This suggests that $\kon_3$ may have the effect of inducing
negative correlations, while $\koff_3$ induces positive
correlations. We also point out the presence of a small peak in
$C_2(\tau)$ in \fig \ref{fig:twoswitches} (indicated by the arrow)
which suggests the presence of a time delay: when one switch flips,
the other tends to follow a short time later. We leave the detailed
properties of these correlations and their parameter dependence to
future work.

\section{Summary and Outlook}
In this paper we have made a detailed study of a generic model of a
binary genetic switch with linear feedback. The model system was
defined in section II by the system of chemical reactions
(\ref{eq:react}).  Linear feedback arises in this switch because the
flipping enzyme $R$ is produced only when the switch is in the on
state, and the rate of flipping to the off state increases linearly
with the amount of $R$. Thus, when the switch is in the on state the
system dynamics inexorably leads to a flip to the off state. We have
shown that this effect can produce a peaked flip time distribution and
a bimodal probability distribution for the copy number of $R$. A mean
field description does not reproduce this phenomenology and so a 
stochastic analysis is required.

We have studied this model analytically, obtaining exact solutions for
the steady state distribution of the number of $R$ molecules, as well
as for the flip time distributions in the two different measurement
ensembles defined in Section \ref{sec:firstpassage}, the Switch Change
Ensemble and the Steady State Ensemble. We have shown how these
ensembles are related and demonstrated that the flip time distribution
in the Switch Change Ensemble may exhibit a peak but the flip time
distribution in the Steady State Ensemble can  never do so.
%An important corollary is
%that to detect the peak experimentally, care must be taken to work in
%the Switch Change Ensemble where measurements are begun when the
%switch has just flipped into the on state. 
We also provide a generic relationship between the flip time
distribution sampled in the two different ensembles. Given that in
single-cell experiments, measuring the flip time distribution in the
SCE is much more demanding than in the SSE, our result provides a way
to access the SCE flip time distribution by making measurements only
in the SSE. Our flip time calculations are reminiscent of persistence
problems in non-equilibrium statistical physics where, for example,
one is interested in the time an Ising spin stays in one state before
flipping.  However, because of the linear feedback of our model
switch, the flip time distribution is not expected to have a long tail
as in usual persistence problems, rather it is the shape of the peak
of the distribution which is of interest.

By studying numerically the time correlations of a single switch,
using the two time autocorrelator (\ref{Ctau}), we have shown that our
model switch can play the role of a primitive ``memory module''. The
two time autocorrelator displays nontrivial behaviour including rather
slow decay, which would be worthy of further study. We have also
investigated the behaviour of two coupled switches within the same
cell, and showed that both positive and negative correlations could be
produced by choosing the parameters appropriately.  In particular for
$\koff_3=0$, as is the case for the {\em fim} switch,
anti-correlations were observed, implying that if one switch were on
at time $t$, the other would tend to be off at that time and for a
subsequent time of about one switch period.

Many open questions and problems remain. At a technical level one
would like to compute correlations of a single switch analytically and
be able to treat the multiple switch system. The model itself could be
refined in several ways, for example, by introducing nonlinear
feedback\cite{FGM07,PB08}. It has been shown that such feedback allows
nontrivial behaviour even at the level of a piecewise deterministic
Markov process approximation \cite{PB08}, where one assumes a
deterministic evolution for the enzyme concentration, but a stochastic
description for the switching. At present our model includes no
explicit coupling to the environment, but such coupling could be
included in a simple way by adding into the model environmental
control of parameters $k_3$ or $k_4$. To make a closer connection to
real biological switches, such as {\em{fim}}, one could extend the
model to include, for example, multiple and cooperative binding of the
enzymes \cite{wolf2002,CB07}. One particularly exciting direction,
which we plan to pursue in future work, is to develop models for
growing populations of switching cells, in which cell growth is
coupled to the switch state. Such models could lead to a better
understanding of the role of phase variation in allowing cells to
survive and proliferate in fluctuating environments. \\

\begin{acknowledgments}
 The authors are grateful to Aileen Adiciptaningrum, David Gally and
 Sander Tans for useful discussions. R. J. A. was funded by the Royal
 Society of Edinburgh. This work was supported by EPSRC under grant
 EP/E030173.
\end{acknowledgments}
\appendix

\section{Solution for the survival probability}
\label{app:char}
We show here how to solve \eq (\ref{eq:survon}) using the method of
characteristics (see e.g. \cite{courant}). Introducing the new
variable $r(z,t)$, we set
\begin{multline}
\frac{d\hont(z(r),t(r))}{dr} = \frac{\p t}{\p r} \frac{\p}{\p t}
\hont(z,t) + \frac{\p z}{\p r} \frac{\p}{\p z} \hont(z,t)\\
=\frac{\p}{\p t} \hont(z,t) + \left[k_1(z-1) + \kon_3
z\right]\frac{\p}{\p z} \hont(z,t) \,\,.
\end{multline}
We can then identify the derivatives of $t$ and $z$ with respect to
$r$ as:
\begin{equation}
  \label{eq:chareq}
  \frac{d t}{d r}=1\,\,,\qquad \frac{d z}{d r}=k_1(z-1) + \kon_3 z
\,\,.
\end{equation}
Next, we solve these equations for $t(r)$ and $z(r)$ using initial
conditions $t(0)=0$ and $z(0)=z_0$:
\begin{equation}
  \label{eq:charsol}
  t(r)=r\,\,,\qquad %z=-\frac{k_1}{k_1+\kon_3} + \left[z_0 +
%  \frac{k_1}{k_1+\kon_3}\right]e^{(k_1+\kon_3)r}\,\,,
 z(r)=k_1 \tau_\on + e^{r/\tau_\on} (z_0 - k_1 \tau_\on ) \,\,,
\end{equation}
where $\tau_\on=(k_1 + \kon_3)^{-1}$. The reduced ordinary
differential equation (ODE) for $\hont$ is:
\begin{equation}
  \frac{d \hont(r)}{d r}=[k_2 (z(r)-1) - \kon_4] \hont(r)\,\,,
\end{equation}
Substituting in the above relation $z(r)$ with its expression given in
(\ref{eq:charsol}), we get an ordinary differential equation for
$\hont(r)$, which can be solved by separation of variables:
\begin{equation}
\frac{d \hont}{\hont}= \tau_\on (-k_2 \kon_3 - \kon_4/\tau_\on +
e^{r/\tau_\on} k_2 (z_0/\tau_\on - k_1)) \,dr \,\,.
\end{equation}
Solving the above equation using the
initial condition $\hont(r=0)=\tW(z_0)$, we arrive at 
\begin{multline}
\hont(r)= \exp \bigg[- \omega_\on r + k_1 k_2 \tau_\on^2 \\- k_2
\tau_\on (e^{r/\tau_\on} (k_1 \tau_\on - z_0) + z_0) \bigg]
\tW(z_0)\,\,,
\end{multline}
where $\omega_\on= \kon_4 + k_2 (1- k_1 \tau_\on)$. Substituting then
from (\ref{eq:charsol}) $r \to t$ and $z_0 \to k_1 \tau_\on +
e^{-t/\tau_\on} (z-k_1 \tau_\on)$ one finally recovers (\ref{eq:hton}).
%and looking at $t$
%and $z$ as functions of $r$, one gets the following characteristic
%equations for (\ref{eq:survon}):
%\begin{equation}
%  \label{eq:chareq}
%  \frac{d t}{d r}=-1\,\,,\qquad \frac{d z}{d r}=k_1 - (k_1 + \kon_3)
%  z\,\,.
%\end{equation}
%Taking the initial conditions $t(0)=0$ and $z(0)=z_0$, the solution to
%the characteristic equations is given by:
%\begin{equation}
%  \label{eq:charsol}
%  t=-r\,\,,\qquad z=k_1 \tau_\on + e^{-r/\tau_\on} (z_0 - k_1
%  \tau_\on)\,\,,
%\end{equation}
%where $\tau_\on=1/(k_1 + \kon_3)$. Then, the reduced ordinary
%differential equation (ODE) for $\hont$ is:
%\begin{equation}
%  \frac{d \hont(r)}{d r}=[k_2 z(r) - (k_2 + \kon_4)] \hont(r)\,\,,
%\end{equation}
%with $z(r)$ given by (\ref{eq:charsol}). Solving the above ODE with
%initial condition $\hont(r)=\tW(z_0)$ yields the final result for
%$\hont$. Substituting $z_0$ and $r$ using (\ref{eq:charsol}), one
%finally recovers (\ref{eq:hton}).

\section{Backwards Evolution Equations for Flip Time Distribution}
\label{sec:bev}
In this appendix we show how the result (\ref{eq:linkbis}) can be
obtained by considering the {\em backward survival probability}:
\begin{equation}
h_s^-(n_0,t) = h_s(n,0|n_0,-t)\,\,,
\end{equation}
which is the probability that the system has survived in the state $s$
without flipping and with $n$ enzymes at time 0 knowing that it had
$n_0$ enzyme molecules at a past time $-t$. The probability $h^-_s$
will verify the backward master equation
\begin{multline}
\label{eq:backwardme}
\frac{\p}{\p t} h_s^-(n_0,t)= n_0 k_1 h_s^-(n_0-1,t) + k_2^s
h_s^-(n_0+1,t) \\- (n_0 k_1 + k^s_2 + n_0 k_3^s + k_4^s)
h_s^-(n_0,t)\,\,.
\end{multline}

In section \ref{sec:firstpassage} we used the forward master equation
to compute the flip time distribution in two steps. First, we computed
the forward survival probability $h_s(n,t)$ with two possible initial
conditions, to distinguish the two possible scenarios of
measurement. Second, we summed this survival probability over all
possible final configurations, and took the time derivative in order
to enforce a flipping at the end of the sampling.

An analogous calculation (which we do not detail) can be carried out
considering the backward master equation (\ref{eq:backwardme}), and
the final result has to be the same. In fact, we can consider the
r.h.s. of (\ref{eq:backwardme}) as a generator of the backward
dynamics. Thus the solution of the backward evolution equation will
have as boundary condition the statistics of the final configuration
at time 0, and will yield the statistics of the possible corresponding
initial configurations at $-t$ (with the additional constraint that
the switch never flipped). Since for both SCE and SSE we condition
that on switch flips at $t=0$, the boundary condition of
(\ref{eq:backwardme}) has to be taken when the switch is flipping from
state $s$ to state $1-s$, and thus corresponds to:
\begin{equation}
h_s^-(n,0)=W^\sce_{1-s}(n)\,\,,
\end{equation}
where $W^\sce_s$ is defined in (\ref{eq:w1}). This is the analogue of
the first step described above. The advantage is that now our boundary
condition is the same for both the SCE and the SSE.

We can relate $h_s^-$ to $F_s$ by noting that $\sum_{n_0}
h_s^-(n_0,t)$ is the probability that the switch has not flipped going
backward for a time $t$. We now have to made a distinction between the
SCE and the SSE, since what happens at time $-t$ is precisely the
initial ensemble. For the case of the SCE, we want the switch to flip
at time $-t$, therefore the flip time distribution is given by:
\begin{equation}
\label{eq:scebis}
F_s^\sce(T)= - \p_T \sum_{n_0} h^-_s(n_0, T)\,\,.
\end{equation}
On the other hand, for the case of the SSE, there is no flipping at
$-t$ to enforce and the flip time distribution $F_s^\sse$ is simply
proportional to the survival probability:
\begin{equation}
\label{eq:ssebis}
F_s^\sse(T)=\frac{ \sum_{n_0} h^-_s(n_0, T)}{\int_0^\infty d T'
\sum_{n_0} h^-_s(n_0, T')}\,\,.
\end{equation}
The denominator in (\ref{eq:ssebis}) is chosen to ensure normalisation
$\int dT F_s^\sse(T) =1$.

Furthermore, we can compute the average flip time in the SCE using
(\ref{eq:scebis}):
\begin{multline}
\label{eq:furthermore}
\avg{T}^\sce_s=\int_0^\infty dT' \,T' F^\sce(T') \\
%= - \int dT' \,T' \,\p_{T'} h(1,T')}{d T'} 
= \int_0^\infty d T' \sum_{n_0} h^-_s(n_0, T')\,\,,
\end{multline}
where an integration by parts has been performed. We can see then that
the denominator in \eq (\ref{eq:ssebis}) is exactly the average
flip time. Finally, integrating \eq (\ref{eq:scebis}) from $T$ 
to infinity and replacing the result in
(\ref{eq:furthermore}), we obtain
\begin{equation}
F^\sse_s(T)=\frac{\int_{T}^\infty F^\sce_s(T')
\,\,dT'}{\int_0^\infty T' F^\sce_s(T') \,\,dT'}\,\,.
\end{equation}
and the result (\ref{eq:linkbis}) is recovered.

%\bibliography{refs}

\end{document}